\documentstyle[11pt]{article}

\newtheorem{conjecture}{Conjecture}

\def\beq{\begin{equation}} \def\eeq{\end{equation}}
\def\bea{\begin{eqnarray}} \def\eea{\end{eqnarray}}
\let\nn=\nonumber
\def\beann{\begin{eqnarray*}} \def\eeann{\end{eqnarray*}}

 \let\k=\kappa \let\la=\lambda 
  \let\p=\pi \let\r=\rho

\def\0{\over } \def\1{\vec }     \def\2{{1\over2}} \def\4{{1\over4}}
\def\5{\bar }  \def\6{\partial } \def\7#1{{#1}\llap{/}}

\def\<{\langle } \def\>{\rangle }

\newfont{\fettohne}{cmssbx10 scaled 1000}
\newfont{\gfettohne}{cmssbx10 scaled 1200}
\newfont{\kfettohne}{cmssbx10 scaled 700}

\renewcommand{\theequation}{\arabic{section}.\arabic{equation}}

\begin{document}
\baselineskip 19pt
\parskip 7pt
\thispagestyle{empty}


\vspace{24pt}

\begin{center}
{\large\bf 
Algebraic Solution of the Hubbard Model on the Infinite 
Interval
}

\vspace{24pt}

Shuichi {\sc Murakami}\ ${}^{1}$
\footnote[3]{E-mail:{\tt murakami@appi.t.u-tokyo.ac.jp}} 
and Frank {\sc G\"ohmann}\ ${}^{2}$
\footnote[4]{E-mail:{\tt Frank.Goehmann@theo.phy.uni-bayreuth.de}}
\footnote[5]{Address from September 1, 1997: 
Institute for Theoretical Physics, State University of New York at Stony 
Brook, Stony Brook, NY 11794-3840, USA}

\vspace{8pt}
${ }^{1}$  {\sl Department 
of Applied Physics, Faculty of Engineering,} \\
{\sl University of Tokyo,} \\
{\sl Hongo 7-3-1, Bunkyo-ku, Tokyo 113, Japan.} \\
${ }^{2}$ {\sl Physikalisches Institut der Universit\"at Bayreuth, TP1,} \\
{\sl  95440 Bayreuth, Germany.}

(Received:\makebox[15em]{} 

\end{center}

\vspace{6pt}

\begin{center}
{\bf Abstract}
\end{center}

We develop the quantum inverse scattering method for the
one-dimensional Hubbard model on the infinite line at zero density.
This enables us to diagonalize the Hamiltonian algebraically. The
eigenstates can be classified as scattering states of particles,
bound pairs of particles and bound states of pairs. We obtain the
corresponding creation and annihilation operators and calculate
the $S$-matrix. The Hamiltonian on the infinite line is invariant
under the Yangian quantum group Y(su(2)). We show that the
$n$-particle scattering states transform like $n$-fold tensor
products of fundamental representations of Y(su(2)) and that the
bound states are Yangian singlet.




\newpage

\section{Introduction}

The past two decades have seen a rapid development of algebraic 
methods for the exact solution of one-dimensional quantum systems. 
This development was partly initiated and most strongly influenced by 
the contribution of the Leningrad school, which clarified the 
fundamental meaning of the Yang-Baxter equation for the 
understanding of exactly solvable one-dimensional 
systems~\cite{TaFa79,KuSk82a,KuSk82b,Ji90}. 
It culminated in the invention of quantum 
groups~\cite{KBIbook} which are by now generally accepted as the 
mathematical framework of the theory.

 The more traditional (coordinate) Bethe ansatz approach to
one-dimensional quantum systems consists of a direct construction
of eigenfunctions and yields a system of Bethe ansatz equations
for a set of parameters which characterize these eigenfunctions
and the spectrum of the Hamiltonian. One may ask, if the quantum
inverse scattering method is more than an alternative way to derive
the Bethe ansatz equations. In fact, as long as we are only
interested in quantities, which are entirely determined by the
spectrum of the Hamiltonian, we do not need the quantum inverse
scattering method. The most successful attempts on the calculation
of correlation functions~\cite{SmiBOOK,JiMiBOOK}, 
however, rely heavily on the use of algebraic methods.

Unfortunately, some of the physically more interesting models, like 
the Hubbard model~\cite{Hubbardbook}, the Kondo model~\cite{TsWi,AnFuLo} or 
the Anderson model~\cite{TsWi} have only partly been capable 
by an algebraic treatment so far. In the present article we will report 
on some recent progress concerning the Hubbard model. 

The Hubbard model was solved by (coordinate) Bethe ansatz by Lieb and 
Wu~\cite{LiWu68,Ya67,Ga67}. 
The basic tools of the quantum inverse scattering method 
for the system under periodic boundary conditions, $R$-matrix and 
monodromy matrix, were constructed by Shastry~\cite{Sha86a,Sha86b,Sha88} 
and by Olmedilla et al.~\cite{WaOlAk87,OlWaAk87,OlWa88}.
Since there exists a so-called pseudo vacuum state, on which the 
monodromy matrix is acting tridiagonally and which is an eigenstate of its 
diagonal elements, an algebraic construction of eigenstates, 
usually called algebraic Bethe ansatz, should be possible 
according to common belief. However, algebraic Bethe ansatz, if we 
understand it in this broad sense, is not really a method. It merely 
means to use commutation relations between the matrix elements of the 
monodromy matrix, whose interpretation in physical terms is moreover 
not a priori clear, to construct the eigenstates of some suitably 
chosen generating function of a family of commuting operators, 
which contains the Hamiltonian of the system. The most 
sophisticated variant of algebraic Bethe ansatz was developed by 
Tarasov~\cite{Ta}, who diagonalized the transfer matrix of the 
Izergin-Korepin model~\cite{IzKo81a}. This work was recently
generalized to the case of the Hubbard model in a remarkable 
article of Ramos and Martins~\cite{RaMa96}. 
Besides the fact that their proof of the ``cancellation of 
unwanted terms" seems to be incomplete, we still feel  
unsatisfied about two points. 
First, the expressions for the eigenstates are of complicated 
recursive nature. It seems to be unlikely that they can be 
used in the calculation of correlation functions. 
Second, due to the complicated way in which the various operators 
contained in the monodromy matrix enter the expression for the
eigenstates, an intuitive physical interpretation is difficult. 

Therefore we follow a different route here~\cite{MuGo97} 
which is based on the work of one of us on the fermionic 
nonlinear Schr{\"o}dinger model~\cite{MuWa96a,MuWa96b}. 
We take the thermodynamic limit first and construct the eigenstates 
afterwards. This is the original idea of the quantum inverse 
scattering method~\cite{Fa80,Tha81,Sk}, which was designed in close 
analogy to the inverse scattering theory for the solution of
classical integrable systems~\cite{FaTa}. The disadvantage of this 
method is that it is so far restricted to uncorrelated vacua 
(ground states). It is therefore neither capable of relativistic models 
nor of non-relativistic models with general nonzero density of 
particles in the ground state. On the other hand, there are lots of 
advantages which make it highly desirable to generalize the method. 
We get rid of the complicated Bethe ansatz equations. 
The algebra of the elements of the monodromy matrix simplifies, and 
we obtain an intuitive interpretation of these operators. 
Furthermore, there is a realistic perspective to calculate correlation 
functions by use of the quantum Gelfand-Levitan equation~\cite{CrThWi80}. 

In the following section we summarize several results for the finite 
periodic system~\cite{WaOlAk87,GoMu97}, which will be needed later. In
section 3 we describe the passage to the infinite interval. 
Some of the technical details are shifted to \ref{appendix:singular}.
We obtain a renormalized monodromy matrix and the commutation 
relations among its entries, which are encoded in a new 
simplified $R$-matrix. In section  4 we identify a suitable generating 
function of commuting operators, namely the quantum determinant 
of a submatrix $A(\lambda)$ of the monodromy matrix. 
The commutation relations between the entries of this submatrix 
decouple from the rest of the algebra and can be written in form 
of an exchange relation which is generated by a submatrix 
$r(\lambda,\mu)$ of the new $R$-matrix. $r(\lambda,\mu)$ is $4\times 4$
and after an appropriate reparametrization turns into the 
rational $R$-matrix of the isotropic Heisenberg spin chain. Thus 
$A(\lambda)$ provides a representation of the Yangian quantum group 
Y(su(2)). In section 5 we identify some of the elements of the 
monodromy matrix as creation and annihilation operators of 
eigenstates of the Hubbard Hamiltonian. We discuss the structure of 
higher conserved quantities, and we calculate the action of the 
Yangian on the eigenstates. In the first part of section 6 we
propose two pairs of normalized creation and annihilation operators of
scattering states. These operators constitute representations of the 
right and left Zamolodchikov-Faddeev algebra, respectively, and 
mutually anticommute. We interpret them as generators of 
fermionic quasi-particles. The Zamolodchikov-Faddeev 
algebra provides their $S$-matrix.
These operators create the known Bethe ansatz states on the infinite 
interval. We work out the action of the Yangian on the 
Bethe ansatz states. The Yangian mixes spin multiplets, which are 
degenerate in the thermodynamic limit. In the second part of section 6 
we propose creation and annihilation operators of bound states of 
quasi-particles. These bound states correspond to the 
string states of coordinate Bethe ansatz~\cite{Ta72}. We calculate their 
$S$-matrix. All of them are Yangian singlet. Section 7 is 
left for a summary and outlook. Appendix B contains a list of 
all commutation relations between the elements of the monodromy 
matrix. Appendix C is devoted to a discussion of the construction of 
bound state operators.
In Appendix D we give explicit expressions for some higher
conserved operators, which are needed for the discussion in section 5.

\section{Hamiltonian and Monodromy Matrix under Periodic Boundary Conditions}
\setcounter{equation}{0}

The Hubbard model in its most basic single-band version is describing 
the dynamics of interacting electrons inside the conduction band 
of a solid. Its Hamiltonian is usually formulated in terms of creation 
and annihilation operators $c_{j\sigma}^{\dagger}, \ c_{j\sigma}$ 
of electrons in Wannier states,
\begin{equation}
\hat{H}=-\sum_{j,\sigma=\uparrow,\downarrow}
(c_{j+1,\sigma}^{\dagger}c_{j,\sigma}+
c_{j,\sigma}^{\dagger}c_{j+1,\sigma})+
U\sum_{j}\left[
\left( n_{j\uparrow}-\frac{1}{2}\right)
\left( n_{j\downarrow}-\frac{1}{2}\right)
-\frac{1}{4}\right].
\label{eqn:Hamiltonian}
\end{equation}
The index $j$ runs over all Wannier states which may be identified with 
the lattice sites of the solid.
$\sigma=\uparrow,\downarrow$ is the spin index and 
$n_{j\sigma}=c_{j\sigma}^{\dagger}c_{j\sigma}$ is the density operator. 
The interaction part of the Hamiltonian is thought to model the screened 
Coulomb interaction of the electrons.
In the following we will use the terminology of quantum field theory, 
and we will call the empty band state $|0\rangle$ the zero density vacuum.
Since we want to study finite excitations over the zero density vacuum 
$|0\rangle$ in the thermodynamic limit, we normalized the Hamiltonian 
such that $\hat{H}|0\rangle=0$.

The most distinctive feature of the one-dimensional model is that 
its Hamiltonian may be embedded into a family of mutually commuting 
operators, which is generated by a properly constructed transfer 
matrix. This feature is commonly referred to as quantum integrability. 
Below we will give a short account of the construction of the 
transfer matrix and its related monodromy matrix for the Hubbard  
model under periodic boundary conditions. 
We start from the local exchange relation~\cite{OlWaAk87}
\begin{equation}
{\cal R}(\lambda,\mu)[ {\cal L}_{j}(\lambda)
\otimes_{{\rm s}} {\cal L}_{j}(\mu) ]
=[ {\cal L}_{j}(\mu)\otimes_{{\rm s}} {\cal L}_{j}(\lambda) ]
{\cal R}(\lambda,\mu).
\label{eqn:RLL}
\end{equation}
The symbol $\otimes_{{\rm s}}$ in this equation 
denotes the Grassmann direct product
\begin{equation}
[A\otimes_{{\rm s}} B]_{\alpha\gamma,\beta\delta}=
(-1)^{[P(\alpha)+P(\beta)]P(\gamma)} A_{\alpha\beta}B_{\gamma\delta}
\end{equation}
with grading $P(1)=P(4)=0,\; P(2)=P(3)=1$.
The $R$-matrix $\cal{R}$ is a c-number matrix which encodes the 
commutation rules of the matrix elements of the $L$-matrix ${\cal L}_{j}$. 
These matrix elements are operators acting on the $j$-th Wannier 
state.
We adopt the expressions for the matrices ${\cal R}$ and 
${\cal L}_{j}$ in terms of 
two parameterizing functions $\alpha(\lambda), \;\gamma(\lambda)$ 
from ref.~\cite{OlWaAk87}. For later convenience, however, we 
shift the arguments of $\alpha(\lambda)$ and $\gamma(\lambda)$ 
by $\frac{\pi}{4}$, such that we simply have $\alpha(\lambda)=\cos \lambda$, 
$\gamma(\lambda)=\sin \lambda $. Then the $L$-matrix is 
\begin{equation}
{\cal L}_{j}(\lambda)=
\left(
\begin{array}{cccc}
-{\rm e}^{h(\lambda)}f_{j\uparrow}f_{j\downarrow} & 
-f_{j\uparrow}c_{j\downarrow} & {\rm i} c_{j\uparrow}f_{j\downarrow} 
&{\rm i} c_{j\uparrow}c_{j\downarrow} {\rm e}^{h(\lambda)} 
\\
-{\rm i} f_{j\uparrow}c_{j\downarrow}^{\dagger} 
&  {\rm e}^{-h(\lambda)}f_{j\uparrow}g_{j\downarrow} 
&  {\rm e}^{-h(\lambda)}c_{j\uparrow}c_{j\downarrow}^{\dagger} 
& {\rm i} c_{j\uparrow}g_{j\downarrow} 
\\
c_{j\uparrow}^{\dagger}f_{j\downarrow} 
& {\rm e}^{-h(\lambda)}c_{j\uparrow}^{\dagger}c_{j\downarrow}
& {\rm e}^{-h(\lambda)} g_{j\uparrow}f_{j\downarrow} 
& g_{j\uparrow}c_{j\downarrow} 
\\
-{\rm i} {\rm e}^{h(\lambda)}c_{j\uparrow}^{\dagger}
c_{j\downarrow}^{\dagger}
& c_{j\uparrow}^{\dagger}g_{j\downarrow}
& {\rm i} g_{j\uparrow}c_{j\downarrow}^{\dagger}
&-g_{j\uparrow}g_{j\downarrow} {\rm e}^{h(\lambda)}
\end{array}
\right) ,
\end{equation}
where 
$f_{j\sigma}(\lambda)=(1-n_{j\sigma})\sin\lambda+
{\rm i}n_{j\sigma}\cos\lambda$, 
$g_{j\sigma}(\lambda)=(1-n_{j\sigma})\cos\lambda-
{\rm i}n_{j\sigma}\sin\lambda$, and  
$h(\lambda)$ is defined as 
\begin{equation}
\frac{\sinh 2h(\lambda)}{\sin 2\lambda}=\frac{U}{4}.
\label{eqn:U4}
\end{equation}
Due to space limitations we do not reproduce the $R$-matrix here. 
It is $16\times 16$ and contains 36 non-vanishing 
entries, only ten of which are different modulo signs.
The ten different entries are denoted by $\rho_{i}$, $i=1,\cdots,10$, 
in ref.~\cite{OlWaAk87}. 
They are rational functions of $\cos\lambda$, $\sin\lambda$ 
and ${\rm e}^{h(\lambda)}$. A list of the matrix elements and some basic 
formulae which have been used in our calculations can be found 
in Appendix A of ref.~\cite{MuGo97}.
Eq.(\ref{eqn:RLL}) considered as an abstract definition of an 
algebra has the property that tensor products of representation 
of this algebra are again representations. This property is
called co-multiplication property. It assures that the monodromy matrix
\begin{equation}
{\cal T}_{mn}(\lambda)={\cal L}_{m-1}(\lambda){\cal L}_{m-2}
(\lambda)\cdots {\cal L}_{n}(\lambda) \ \  (m>n)
\label{eqn:finiteT}
\end{equation}
satisfies again eq.(\ref{eqn:RLL}). ${\cal T}_{mn}(\la)$ contains 
all information about the Hubbard model under periodic boundary 
conditions. After proper renormalization ${\cal T}_{mn}(\la)$ 
turns into the monodromy matrix of the Hubbard model in the 
thermodynamic limit, which will be the central object of 
investigation of the present paper. Before we continue with the 
description of the thermodynamic limit, we list the most important 
properties of ${\cal T}_{mn}(\lambda)$~\cite{GoMu97}.

The transfer matrix
\begin{equation}
\tau_{mn}(\la)={\rm str}({\cal T}_{mn}(\la))={\rm tr}
((\sigma^{z}\otimes\sigma^{z}){\cal T}_{mn}(\la))
\end{equation}
generates a family of mutually commuting 
operators~\cite{Sha86a,Sha86b,OlWaAk87,GoMu97}, 
\begin{eqnarray}
\ln\tau_{mn}(\la)&=& \frac{{\rm i}\pi}{2}(\hat{N}_{mn}-2(m-n))+
{\rm i}\hat{\Pi}_{mn}\nonumber \\
&&\makebox[1em]{}+\lambda\left( \hat{H}_{mn}+(m-n)\frac{U}{4}\right)
+O(\lambda^{2}).
\label{eqn:expandtau}
\end{eqnarray}
Here 
\begin{equation}
  \hat{N}_{mn}=\sum_{j=n}^{m-1}(n_{j\uparrow}+n_{j\downarrow})
\label{eqn:N}
\end{equation}
is the particle number operator, and $\hat{\Pi}_{mn}$ is the 
lattice momentum operator. For the subtle question how to define 
$\hat{\Pi}_{mn}$ properly such that it commutes with the Hamiltonian 
we refer the reader to Ref.~\cite{GoMu97}.
$\hat{H}_{mn}$ is the Hamiltonian (\ref{eqn:Hamiltonian}) on 
a one-dimensional lattice of length $m-n$ under periodic boundary 
conditions (let $j$ run from $n$ to $m-1$ in eq.(\ref{eqn:Hamiltonian})
and let $c_{m}=c_{n}$).

Clearly, the Hamiltonian (\ref{eqn:Hamiltonian}) is invariant under 
su(2)-rotations generated by the operators of total spin
\begin{equation}
  S^{a}=\frac{1}{2}\sum_{j=n}^{m-1}
\sigma_{\alpha\beta}^{a}c_{j\alpha}^{\dagger}c_{j\beta}.
\label{eqn:spin}
\end{equation}
Here $a=x,y,z$ and the matrices $\sigma^{a}$ are the Pauli matrices. 
Due to the invariance of the Hamiltonian under the 
transformation
\begin{equation}
  \label{eqn:transformation}
  c_{j\uparrow}\rightarrow c_{j\uparrow}, \ \ 
  c_{j\downarrow}\rightarrow (-1)^{j}c_{j\downarrow}^{\dagger}, \ \ 
  U\rightarrow -U
\end{equation}
there exists a second su(2) symmetry~\cite{EKS,HeLi71,Ya89}.
This symmetry is called $\eta$-pairing symmetry.
Applying 
(\ref{eqn:transformation}) to (\ref{eqn:spin}) we get its 
generators $\eta^{a}$ in the form 
\begin{eqnarray}
  \eta^{x}&=&
   \frac{1}{2}\sum_{j=n}^{m-1}(-1)^{j}
    (c_{j\uparrow}^{\dagger}c_{j\downarrow}^{\dagger}+
    c_{j\downarrow}c_{j\uparrow}),\\
  \eta^{y}&=&
  -\frac{{\rm i}}{2}\sum_{j=n}^{m-1}(-1)^{j}
    (c_{j\uparrow}^{\dagger}c_{j\downarrow}^{\dagger}-
    c_{j\downarrow}c_{j\uparrow}),\\
  \eta^{z}&=&
   \frac{1}{2}(\hat{N}_{mn}-m+n).
\end{eqnarray}
Note that the transformation (\ref{eqn:transformation}) twists the boundary 
conditions, if $m-n$ is odd.  Therefore $\hat{H}_{mn}$ commutes with 
$\eta^{x}$ and $\eta^{y}$ only if the lattice has an even number of sites. 
Since $\eta^{z}$ is essentially the particle number operator, we may 
understand the $\eta$-pairing symmetry as a non-Abelian extension of 
gauge symmetry. 

Rotational symmetry and $\eta$-pairing symmetry both extend to symmetries
of the monodromy matrix~\cite{GoMu97}. In order to make this statement
explicit, we have to introduce certain matrix representations of 
su(2). We will denote the $n\times n$ unit matrix by $I_{n}$. Let
\begin{eqnarray}
  \Sigma^{x}_{s}&=&\frac{1}{2}(
   \sigma^{+}\otimes\sigma^{-}+\sigma^{-}\otimes\sigma^{+}), \\
  \Sigma^{y}_{s}&=&-\frac{{\rm i}}{2}(
   \sigma^{+}\otimes\sigma^{-}-\sigma^{-}\otimes\sigma^{+}), \\
  \Sigma^{x}_{s}&=&\frac{1}{4}(
   \sigma^{z}\otimes I_{2}-I_{2}\otimes\sigma^{z}), 
\end{eqnarray}
and
\begin{eqnarray}
  \Sigma^{x}_{\eta}&=&\frac{1}{2}(
   \sigma^{+}\otimes\sigma^{+}+\sigma^{-}\otimes\sigma^{-}), \\
  \Sigma^{y}_{\eta}&=&-\frac{{\rm i}}{2}(
   \sigma^{+}\otimes\sigma^{+}-\sigma^{-}\otimes\sigma^{-}), \\
  \Sigma^{z}_{\eta}&=&\frac{1}{4}(
   \sigma^{z}\otimes I_{2}+I_{2}\otimes\sigma^{z}).
\end{eqnarray}
These matrices obviously satisfy the su(2) commutation rules
\begin{equation}
  [\Sigma_{j}^{a},\Sigma_{j}^{b}]={\rm i}\varepsilon^{abc}
    \Sigma_{j}^{c}, \ j=s,\eta.
\label{eqn:Sigmasu2}
\end{equation}
Let us perform a basis transformation
\begin{equation}
  \tilde{\Sigma}^{a}_{j}=G^{ab}\Sigma^{b}_{j} 
\label{eqn:twist}
\end{equation}
with transformation matrix 
\begin{equation}
  G=\left(
  \begin{array}{ccc}
  0&-1&0 \\
  1&0&0 \\
  0&0&1
  \end{array}
\right).
\label{eqn:SigmaG}
\end{equation}
Since $G$ is orthogonal with ${\rm det}(G)=1$, the transformed matrices 
$\tilde{\Sigma}^{a}_{j}$ satisfy (\ref{eqn:Sigmasu2}). 
We are now able to state the su(2)$\oplus$su(2) symmetry of the 
monodromy matrix ${\cal T}_{mn}(\lambda)$,
\begin{eqnarray}
&&[ {\cal T}_{mn}(\lambda), \tilde{\Sigma}_{s}^{a}+S^{a} ]=0,
\label{eqn:Ts} \\
&&[{\cal T}_{mn}(\lambda),\tilde{\Sigma}_{\eta}^{a}+{\eta}^{a}]=0.
\label{eqn:Teta}
\end{eqnarray}
If $a=x,y$ in (\ref{eqn:Teta}) we have to require 
both $m$ and $n$  
to be odd. (\ref{eqn:Ts}) and (\ref{eqn:Teta}) imply the 
invariance of all higher conserved quantities in the expansion 
(\ref{eqn:expandtau}) under rotations and $\eta$-pairing transformations. 

The twist (\ref{eqn:twist}) may appear somewhat unnatural. However, 
we had to introduce it here, since we wanted to keep the notation 
of the earlier paper~\cite{MuGo97}. 
We may remove the twist by a gauge transformation in auxiliary space. 
Let 
\begin{equation}
  W={\rm diag}(1,1,{\rm i},{\rm i}).
\end{equation}
Then
\begin{equation}
  \tilde{\Sigma}_{j}^{a}=W^{-1}\Sigma_{j}^{a}W.
\end{equation}
Thus the gauge transformed monodromy matrix 
$W{\cal T}_{mn}(\lambda)W^{-1}$ satisfies (\ref{eqn:Ts}) and 
(\ref{eqn:Teta}) with $\tilde{\Sigma}_{j}^{a}$ replaced by 
$\Sigma_{j}^{a}$. Note that $W\otimes W$ commutes with the 
$R$-matrix~\cite{GoMu97}, which implies that the exchange 
relations for ${\cal T}_{mn}(\lambda)$ and $W{\cal T}_{mn}(\lambda)W^{-1}$ 
are the same.

The grading of the monodromy matrix, its behavior under 
particle-hole transformations~\cite{GoMu97} and the structure of the
matrices $\Sigma_{j}^{a}$ suggest the following block notation for 
the monodromy matrix, 
\begin{equation}
  {\cal T}_{mn}(\lambda)=
\left(
\begin{array}{cccc}
D_{11}(\lambda) & C_{11}(\lambda) & C_{12}(\lambda) & D_{12}(\lambda) \\
B_{11}(\lambda) & A_{11}(\lambda) &A_{12}(\lambda) & B_{12}(\lambda) \\
B_{21}(\lambda) & A_{21}(\lambda) & A_{22}(\lambda) & B_{22}(\lambda) \\
D_{21}(\lambda) & C_{21}(\lambda) & C_{22}(\lambda) & D_{22}(\lambda) 
\end{array} 
\right).
\label{eqn:matrixT}
\end{equation}
We will see in the following that many of the algebraic properties 
of the Hubbard model are conveniently expressed in terms of 
the $2\times 2$ submatrices 
$A(\lambda)$, $B(\lambda)$, $C(\lambda)$, $D(\lambda)$. 
As an example let us describe the behavior of 
${\cal T}_{mn}(\lambda)$ under hermitian conjugation, which will be
needed later and which can be obtained by using the methods 
outlined in Ref.~\cite{GoMu97},
\begin{eqnarray}
(A(\lambda))^{\dagger}&=&\sigma^{y}A(\pi /2-\lambda^{*})\sigma^{y}, 
\label{eqn:Adagger}\\
(B(\lambda))^{\dagger}&=&{\rm i}
\sigma^{y}B(\pi /2-\lambda^{*})\sigma^{y}, \label{eqn:Bdagger} \\
(C(\lambda))^{\dagger}&=&{\rm i}
\sigma^{y}C(\pi /2-\lambda^{*})\sigma^{y}, \label{eqn:Cdagger} \\
(D(\lambda))^{\dagger}&=&
\sigma^{y}D(\pi /2-\lambda^{*})\sigma^{y}.\label{eqn:Ddagger}
\end{eqnarray}
The dagger on the lhs of these equations means hermitian conjugation in 
quantum space, and the asterisk on the rhs denotes complex conjugation.
For notational convenience we did not attach labels $m$ and $n$ to 
the submatrices $A(\lambda),\cdots,D(\lambda)$ on the rhs of 
(\ref{eqn:matrixT}). We will keep the same notation below, when we 
discuss the thermodynamic limit.

\section{Passage to the Infinite Interval}
\label{sec:passage}
\setcounter{equation}{0}

As we shall see in the sequel, carrying out the thermodynamic
limit leads to a severe simplification of the $R$-matrix. 
The commutation relations between the entries of the monodromy 
matrix will become simple enough to allow for an identification 
of creation and annihilation operators of quasi-particles, 
generators of conserved quantities and symmetry operators. 
The thermodynamic limit cannot be taken na{\"\i}vely. 
The monodromy matrix requires infrared renormalization, 
which has to be done with respect to a given vacuum 
characterized by macroscopic parameters. These are the 
density of electrons $\rho_{N}$ and the magnetization density 
$\rho_{M}$.  As result of the thermodynamic limit we will obtain the 
finite energy excitations over the chosen vacuum. In contrast to 
the algebraic Bethe ansatz for the finite periodic system 
we will not be able anymore to distinguish between a pseudo-vacuum, 
upon which all eigenstates of the transfer matrix are built 
by the action of creation operators, and the physical vacuum, 
which is the true ground state of the model. 
In general, in the thermodynamic limit both states will be characterized
by different values of $\rho_{M}, \ \rho_{N}$ and thus will be separated 
by an infinite energy difference. 

Infrared renormalization of the monodromy  matrix is done in 
analogy with the inverse scattering method for integrable classical 
systems~\cite{FaTa} by splitting off the asymptotics of the vacuum 
expectation value of the monodromy matrix for $m,-n\rightarrow\infty$, 
which therefore has to be known a priori. 
For this reason the method~\cite{Fa80,Tha81,MuGo97,Sk} 
is so far restricted to (asymptotically) uncorrelated vacua. 
In case of the Hubbard model there are four possible choices, 
the empty band ($\rho_{M}=\rho_{N}=0$), the completely filled 
band ($\rho_{M}=0, \rho_{N}=2$), and the half-filled band with 
all spins up ($\rho_{M}=1,\rho_{N}=1$) or all spins down 
($\rho_{M}=-1,\rho_{N}=1$). In the following we will restrict 
ourselves to the empty band vacuum $|0\rangle$ which is defined 
by
\begin{equation}
c_{j\sigma}|0\rangle=0.
\end{equation}

Our description of the general method closely follows Sklyanin~\cite{Sk}.
Define the Hilbert space ${\cal H}$ of states of ``compact support"
as the space of all finite linear combinations of vectors 
$c_{m_{1}\sigma_{1}}^{\dagger}\cdots c_{m_{n}\sigma_{n}}^{\dagger}
|0\rangle$. Denote the vacuum expectation value of the 
$L$-matrix by 
\begin{equation}
V(\lambda)=\langle 0|{\cal L}_{m}(\lambda)|0\rangle .
\end{equation}
$V(\lambda)$ does not depend on $m$, because of the 
translational invariance of the vacuum.

Let 
\begin{eqnarray}
\tilde{{\cal L}}_{j}(\lambda)&=&V(\lambda)^{-j-1}{\cal L}_{j}(\lambda)
V(\lambda)^{j}, \\
\tilde{{\cal T}}_{mn}(\lambda)&=&V(\lambda)^{-m}{\cal T}_{mn}(\lambda)
V(\lambda)^{n}. 
\label{eqn:tildeT}
\end{eqnarray}
It is easy to see that the limits $\lim_{n\rightarrow -\infty}
\langle x |\tilde{{\cal T}}_{mn}(\lambda)|y\rangle$
and $\lim_{m\rightarrow \infty}
\langle x |\tilde{{\cal T}}_{mn}(\lambda)|y\rangle$
exist for all $|x\rangle,|y\rangle\in {\cal H}$. 
These weak limits determine a pair of operators
\begin{eqnarray}
\tilde{{\cal T}}_{m}^{+}(\lambda)&=&\lim_{n\rightarrow +\infty}
\tilde{{\cal T}}_{nm}(\lambda), \\
\tilde{{\cal T}}_{m}^{-}(\lambda)&=&\lim_{n\rightarrow -\infty}
\tilde{{\cal T}}_{mn}(\lambda), 
\end{eqnarray}
with asymptotics
\begin{equation}
\lim_{m\rightarrow +\infty}
\tilde{{\cal T}}_{m}^{+}(\lambda)=
\lim_{m\rightarrow -\infty}
\tilde{{\cal T}}_{m}^{-}(\lambda)=I_{4}.
\label{eqn:TTI}
\end{equation}
Multiplying (\ref{eqn:finiteT}) from the
left by ${\cal L}_{m}(\lambda)$ or from the right by
${\cal L}_{n-1}(\lambda)$, respectively, we obtain two recursion
relations for ${\cal T}_{mn}(\lambda)$, which imply a pair of
recursion relations for $\tilde{{\cal T}}_{m}^{+}(\lambda)$ and
$\tilde{{\cal T}}_{m}^{-}(\lambda)$. By use of the asymptotic condition
(\ref{eqn:TTI}) these are equivalent to the following pair of Volterra
``integral equations" for  $\tilde{{\cal T}}_{m}^{\pm}(\lambda)$,
\begin{eqnarray}
\tilde{{\cal T}}_{m}^{+}(\lambda)&=&I_{4}+\sum_{j=m+1}^{\infty}
\tilde{{\cal T}}_{j}^{+}(\lambda)\ 
(\tilde{{\cal L}}_{j-1}(\lambda)-I_{4}), \label{eqn:defTmplus}\\
\tilde{{\cal T}}_{m}^{-}(\lambda)&=&I_{4}+\sum_{j=-\infty}^{m-1}
(\tilde{{\cal L}}_{j}(\lambda)-I_{4})\  \tilde{{\cal T}}_{j}^{-}(\lambda).
\label{eqn:defTmminus}
\end{eqnarray}
The above considerations imply the existence of the weak limit
\begin{equation}
\tilde{{\cal T}}(\lambda)=\lim_{m,-n\rightarrow\infty}
\tilde{{\cal T}}_{mn}(\lambda)=
\tilde{{\cal T}}_{m}^{+}(\lambda)
\tilde{{\cal T}}_{m}^{-}(\lambda).
\label{eqn:T}
\end{equation}
$\tilde{{\cal T}}(\lambda)$ is the renormalized monodromy matrix. 
Equation (\ref{eqn:defTmplus}), or (\ref{eqn:defTmminus}) respectively, 
implies the ``integral representation"
\begin{equation}
\tilde{{\cal T}}(\lambda)=I_{4}+\sum_{m}(\tilde{{\cal L}}_{m}(\lambda)
-I_{4})+\sum_{m>n}(\tilde{{\cal L}}_{m}(\lambda)-I_{4})
(\tilde{{\cal L}}_{n}(\lambda)
-I_{4})+\cdots .
\label{eqn:expandtildeT}
\end{equation}
Note that $\langle 0|(\tilde{{\cal L}}_{m}(\lambda)-I_{4})|0\rangle=0$
by construction. 
Hence $\langle 0|\tilde{{\cal T}}(\lambda)|0\rangle=I_{4}$. 

$\tilde{{\cal L}}_{m}(\lambda)$ can be easily calculated. 
We find
\begin{equation}
V(\lambda)=
{\rm diag} 
(-{\rm e}^{h(\lambda)}\sin^{2}\lambda,
{\rm e}^{-h(\lambda)}\cos\lambda\sin\lambda,
{\rm e}^{-h(\lambda)}\cos\lambda\sin\lambda,
-{\rm e}^{h(\lambda)}\cos^{2}\lambda),
\label{eqn:defV}
\end{equation}
and thus
\begin{eqnarray}
& & \tilde{\cal L}_{m}(\lambda) =  V(\lambda)^{-m-1}
{\cal L}_{m}(\lambda)V(\lambda)^{m} \nonumber \\
&& \nonumber \\
& & =
\left(
\begin{array}{cc}
({\rm i}\cot {\lambda} )^{n_{m\uparrow}+n_{m\downarrow}}
&({\rm i}\cot {\lambda} )^{n_{m\uparrow}}c_{m\downarrow}
\frac{{\rm e}^{-h(\lambda)}}{\sin \lambda} {\rm e}^{{\rm i}
mp(\lambda)}
\\
-{\rm i}
({\rm i}\cot {\lambda} )^{n_{m\uparrow}}c^{\dagger}_{m\downarrow}
\frac{{\rm e}^{h(\lambda)}}{\cos \lambda} 
{\rm e}^{-{\rm i}mp(\lambda)}
&({\rm i}\cot {\lambda} )^{n_{m\uparrow}-n_{m\downarrow}}
\\
c^{\dagger}_{m\uparrow}
({\rm i}\cot {\lambda} )^{n_{m\downarrow}}
\frac{{\rm e}^{h(\lambda)}}{\cos \lambda} 
{\rm e}^{-{\rm i}mp(\lambda)}
&c_{m\uparrow}^{\dagger}c_{m\downarrow}\frac{1}{\sin \lambda \cos \lambda}
\\
{\rm i} c_{m\uparrow}^{\dagger}c_{m\downarrow}^{\dagger}
\frac{1}{\cos^{2} \lambda}
\tan^{2m} \lambda 
&-c_{m\uparrow}^{\dagger}({\rm i}\cot \lambda)^{-n_{m\downarrow}}
\frac{{\rm e}^{-h(\lambda)}}{\cos \lambda} 
{\rm e}^{-{\rm i}mk(\lambda)}
\end{array}
\right.  \nonumber \\
& & \makebox[5em]{}
\left.
\begin{array}{cc}
-{\rm i}c_{m\uparrow}({\rm i}\cot \lambda)^{n_{m\downarrow}}
\frac{{\rm e}^{-h(\lambda)}}{\sin \lambda} 
{\rm e}^{{\rm i}mp(\lambda)}
& -{\rm i} c_{m\uparrow}c_{m\downarrow}\frac{1}{\sin^{2} \lambda}
\cot^{2m} \lambda 
\\
c_{m\uparrow}c_{m\downarrow}^{\dagger}\frac{1}{\sin \lambda \cos \lambda}
&{\rm i}c_{m\uparrow}({\rm i}\cot \lambda)^{-n_{m\downarrow}}
\frac{{\rm e}^{h(\lambda)}}{\sin \lambda} 
{\rm e}^{{\rm i}mk(\lambda)}
\\
({\rm i}\cot {\lambda} )^{-n_{m\uparrow}+n_{m\downarrow}}
&({\rm i}\cot \lambda)^{-n_{m\uparrow}}c_{m\downarrow}
\frac{{\rm e}^{h(\lambda)}}{\sin \lambda} 
{\rm e}^{{\rm i}mk(\lambda)}
\\
-{\rm i}({\rm i}\cot {\lambda} )^{-n_{m\uparrow}}
c_{m\downarrow}^{\dagger}
\frac{{\rm e}^{-h(\lambda)}}{\cos \lambda} 
{\rm e}^{-{\rm i}mk(\lambda)}
&({\rm i}\cot {\lambda} )^{-n_{m\uparrow}-n_{m\downarrow}}
\end{array}
\right). \label{eqn:tildeL}\makebox[2em]{}
\end{eqnarray}
Here we have introduced new functions 
\begin{equation}
{\rm e}^{{\rm i}k(\lambda)}=
-{\rm e}^{2h(\lambda)}\cot \lambda, 
\makebox[1em]{}
{\rm e}^{{\rm i}p(\lambda)}=
-{\rm e}^{-2h(\lambda)}\cot \lambda, 
\label{eqn:defkp}
\end{equation}
which we adopted from the recent analytic Bethe Ansatz for the Hubbard 
model by Yue and Deguchi~\cite{YuDe96}.

Now we will turn to the calculation of the commutation relations between 
the elements of $\tilde{{\cal T}}(\lambda)$. Let 
\begin{eqnarray}
{\cal L}_{m}^{(2)}(\lambda,\mu)&=&{\cal L}_{m}(\lambda)\otimes_{{\rm s}}
{\cal L}_{m}(\mu), \\
{\cal T}_{mn}^{(2)}(\lambda,\mu)&=&{\cal T}_{mn}(\lambda)\otimes_{{\rm s}}
{\cal T}_{mn}(\mu).
\end{eqnarray}
We may apply the 
above discussion of the renormalization of ${\cal T}_{mn}(\lambda)$ 
to ${\cal T}_{mn}^{(2)}(\lambda,\mu)$, if we replace $V(\lambda)$ by
\begin{equation}
V^{(2)}(\lambda,\mu)=\langle 0|{\cal L}_{m}(\lambda)\otimes_{{\rm s}}
{\cal L}_{m}(\mu)|0\rangle . 
\end{equation}
Note that $V^{(2)}(\lambda,\mu)$ is \underline{not} just the 
tensor product $V(\lambda)\otimes_{{\rm s}}V(\mu)$. 
There appear additional off-diagonal terms due to normal ordering of the 
operators. 
We obtain a renormalized tensor product matrix
\begin{equation}
\tilde{{\cal T}}^{(2)}(\lambda,\mu)=\lim_{m,-n\rightarrow\infty}
V^{(2)}(\lambda,\mu)^{-m}
{\cal T}_{mn}^{(2)}(\lambda,\mu)V^{(2)}(\lambda,\mu)^{n},
\end{equation}
which satisfies $\langle 0|\tilde{{\cal T}}^{(2)}
(\lambda,\mu)|0\rangle=I_{16}$.
Taking the vacuum expectation value of the local exchange relation
(\ref{eqn:RLL}) yields
\begin{equation}
{\cal R}(\lambda,\mu)V^{(2)}(\lambda,\mu)=
V^{(2)}(\mu,\lambda)
{\cal R}(\lambda,\mu),
\end{equation}
and we conclude that 
\begin{equation}
{\cal R}(\lambda,\mu)\tilde{{\cal T}}^{(2)}(\lambda,\mu)=
\tilde{{\cal T}}^{(2)}(\mu,\lambda)
{\cal R}(\lambda,\mu).
\label{eqn:exchangeT2}
\end{equation}
If $\tilde{{\cal T}}_{mn}^{(2)}(\lambda,\mu)$ is defined in analogy to 
$\tilde{{\cal T}}_{mn}(\lambda)$
with $V(\lambda)$ replaced by $V^{(2)}(\lambda,\mu)$ in the 
definition (\ref{eqn:tildeT}), then 
\begin{equation}
\tilde{{\cal T}}_{mn}(\lambda)\otimes_{{\rm s}}
\tilde{{\cal T}}_{mn}(\mu)=
U_{m}(\lambda,\mu)^{-1}
\tilde{{\cal T}}_{mn}^{(2)}(\lambda,\mu)
U_{n}(\lambda,\mu),  
\label{eqn:TTT2}
\end{equation}
where we have set
\begin{equation}
U_{n}(\lambda,\mu)=V^{(2)}(\lambda,\mu)^{-n}[V(\lambda)^{n} 
\otimes_{{\rm s}}V(\mu)^{n}].
\label{eqn:defU}
\end{equation}
Assume for a while that the limits
\begin{equation}
U_{+}(\lambda,\mu)^{-1}=\lim_{m\rightarrow\infty}U_{m}
(\lambda,\mu)^{-1}, \ \ 
U_{-}(\lambda,\mu)=\lim_{m\rightarrow -\infty}U_{m}
(\lambda,\mu).
\label{eqn:defUpm}
\end{equation}
exist in a common domain of convergence. 
Then, according to eq.~(\ref{eqn:TTT2})
$\tilde{{\cal T}}_{mn}(\lambda)\otimes_{{\rm s}}
\tilde{{\cal T}}_{mn}(\mu)$ 
has a weak limit for $m,-n\rightarrow\infty$. 
We identify this limit with $\tilde{{\cal T}}(\lambda)\otimes_{{\rm s}}
\tilde{{\cal T}}(\mu)$, 
\begin{equation}
\tilde{{\cal T}}(\lambda)\otimes_{{\rm s}}
\tilde{{\cal T}}(\mu)=
U_{+}(\lambda,\mu)^{-1}
\tilde{{\cal T}}^{(2)}(\lambda,\mu)
U_{-}(\lambda,\mu).
\label{eqn:TTUTU}
\end{equation}
Finally, inserting the above equation into (\ref{eqn:exchangeT2}), 
we arrive at the exchange relation for the monodromy matrix 
$\tilde{{\cal T}}(\lambda)$ on the infinite interval, 
\begin{equation}
\tilde{{\cal R}}^{(+)}(\lambda,\mu)\left[
\tilde{\cal T}(\lambda)\otimes_{{\rm s}}\tilde{\cal T}(\mu)\right]
=\left[\tilde{\cal T}(\mu)
\otimes_{{\rm s}}\tilde{\cal T}(\lambda)\right]
\tilde{{\cal R}}^{(-)}(\lambda,\mu),
\label{eqn:newRTT}
\end{equation}
where
\begin{equation}
\tilde{\cal R}^{(\pm)}(\lambda,\mu)=
U_{\pm}(\mu,\lambda)^{-1}{\cal R}(\lambda,\mu)
U_{\pm}(\lambda,\mu). 
\label{eqn:tildeRpm}
\end{equation}
The calculation of the matrices $U_{\pm}(\lambda,\mu)$ is rather 
technical. We present it in \ref{appendix:singular}. 
Here we only note three important 
facts. (i) there is no common domain of convergence for all 
matrix elements of $U_{+}(\lambda,\mu)$ and 
$U_{-}(\lambda,\mu)$. We will come back to this point later. 
(ii) if we stay away from some singular points
(cf.\ \ref{appendix:singular}) then
$U_{+}(\lambda,\mu)_{\alpha\beta,\gamma\delta}
= U_{-}(\lambda,\mu)_{\alpha\beta,\gamma\delta}$.
(iii) it is a nontrivial matter of fact that all the matrix elements 
of $U_{\pm} (\lambda, \mu)$ are simple rational functions of the
original Boltzmann weights $\rho_{j} (\lambda, \mu)$.

Using the explicit form of the matrices $U_{\pm}(\lambda,\mu)$ 
given in \ref{appendix:singular}, we obtain
\begin{eqnarray}
\lefteqn{\tilde{\cal R}(\lambda,\mu) =
\tilde{\cal R}^{(+)}(\lambda,\mu)= 
\tilde{\cal R}^{(-)}(\lambda,\mu) = }\nonumber \\[1ex]
& & \hspace{-22pt} \left( 
{\arraycolsep 2pt
\begin{array}{cccccccccccccccc}
\rho_{1} & 0 & 0 & 0 & 
0 & 0 & 0 & 0 & 0 & 0 & 0 & 0 & 0 & 0 & 0 & 0 \\
0 & 0 & 0 & 0 & 
\frac{\rho_{1}\rho_{4}}{{\rm i}\rho_{10}} 
& 0 & 0 & 0 & 0 & 0 & 0 & 0 
& 0 & 0 & 0 & 0 \\
0 & 0 & 0 & 0 & 
0 & 0 & 0 & 0 &
\frac{\rho_{1}\rho_{4}}{{\rm i}\rho_{10}} 
& 0 & 0 & 0 & 0 & 0 & 0 & 0 \\
0 & 0 & 0 & 0 & 
0 & 0 & 0 & 0 &
0 & 0 & 0 & 0 &
\frac{-\rho_{1}\rho_{4}}{\rho_{5}-\rho_{4}} & 0 & 0 & 0 \\
0 & 
 -{\rm i}\rho_{10} &0 & 0 & 0 &  0 & 0 & 0 &
0 & 0 & 0 & 0 &
0 & 0 & 0 & 0 \\
0 & 0 & 0 & 0 & 
0 &  \rho_{4} & 0 & 0 &
0 & 0 & 0 & 0 &
0 & 0 & 0 & 0 \\
0 & 0 & 0 & 0 & 
0 & 0 & \frac{\rho_{3}\rho_{4}-\rho_{2}^{2}}{\rho_{3}-\rho_{1}}
& 0 & 0 & 
\frac{\rho_{9}\rho_{10}}{\rho_{3}-\rho_{1}} & 0 & 0 &
0 & 0 & 0 & 0 \\
0 & 0 & 0 & 0 & 
0 & 0 & 0 & 0 &
0 & 0 & 0 & 0 &
0 & 
\frac{{\rm i}\rho_{1}\rho_{4}}{\rho_{9}}
 & 0 & 0 \\
0 & 0 &  -{\rm i}\rho_{10} & 0 & 
0 & 0 & 0 & 0 &
0 & 0 & 0 & 0 &
0 & 0 & 0 & 0 \\
0 & 0 & 0 & 0 & 
0 & 0 &\frac{\rho_{9}\rho_{10}}{\rho_{3}-\rho_{1}}
& 0 & 0 &  \frac{\rho_{3}\rho_{4}-\rho_{2}^{2}}{\rho_{3}-\rho_{1}}
 & 0 & 0 & 
0 & 0 & 0 & 0 \\
0 & 0 & 0 & 0 & 
0 & 0 & 0 & 0 &
0 & 0 &  \rho_{4} & 0 &
0 & 0 & 0 & 0 \\
0 & 0 & 0 & 0 & 
0 & 0 & 0 & 0 &
0 & 0 & 0 & 0 &
0 & 0 & \frac{{\rm i}\rho_{1}\rho_{4}}{\rho_{9}}
& 0 \\
0 & 0 & 0 &  \rho_{1}-\rho_{3} & 
0 & 0 & 0 & 0 &
0 & 0 & 0 & 0 &
0 & 0 & 0 & 0 \\
0 & 0 & 0 & 0 & 
0 & 0 & 0 &  {\rm i} \rho_{9} &
0 & 0 & 0 & 0 &
0 & 0 & 0 & 0 \\
0 & 0 & 0 & 0 & 
0 & 0 & 0 & 0 &
0 & 0 & 0 &  {\rm i} \rho_{9} &
0 & 0 & 0 & 0 \\
0 & 0 & 0 & 0 & 
0 & 0 & 0 & 0 &
0 & 0 & 0 & 0 &
0 & 0 & 0 &  \rho_{1} 
\end{array}
}
\right)
\raisebox{-22ex}{.} \makebox[2em]{}
\label{eqn:deftildeR}
\end{eqnarray}
The reader is urged to compare this expression with the $R$-matrix on the 
finite interval~\cite{OlWaAk87}. Instead of the 
36 non-vanishing elements of 
the original $R$-matrix we have only 18 non-vanishing elements here, 
which brings about simpler commutation relations between
the elements of the monodromy matrix.  
All matrix elements except the two diagonal elements
$\frac{\rho_{3}\rho_{4}-\rho_{2}^{2}}{\rho_{3}-\rho_{1}}$
are just at the position of the 1's of the permutation matrix, 
which means that, were it not for the two elements
$\frac{\rho_{3}\rho_{4}-\rho_{2}^{2}}{\rho_{3}-\rho_{1}}$,
all commutation relations would reduce to the mere interchange of 
two factors along with a multiplication by some rational 
function of the Boltzmann weights. 
A close investigation of (\ref{eqn:newRTT}) shows in particular, that 
the elements of the submatrices $A(\lambda),\cdots,D(\lambda)$ 
generate sub-algebras of the exchange relation. The sub-algebras 
generated by $A(\lambda)$ and by $D(\lambda)$ are again 
Yang-Baxter algebras with certain new $R$-matrices which
are submatrices of $\tilde{{\cal R}}(\lambda,\mu)$. 
The sub-algebras generated by $B(\lambda)$ and $C(\lambda)$ 
can, after certain normalizations, be identified as 
representations of the right and left Zamolodchikov-Faddeev 
algebra. The construction of these algebras and the discussion 
of their physical meaning will be the contents of the 
following sections.

Let us come back to the remark (i) above. 
Due to the peculiar convergence properties of the limits 
$U_{\pm}(\lambda,\mu)$ the exchange relation in the form 
(\ref{eqn:newRTT}) has only symbolical meaning. 
It has to be understood as generating a set of 256 equations, 
each of which is meaningful. These equations, however, are 
not necessarily defined on the same domain in the $\lambda,\mu$
parameter space (cf.\ \ref{appendix:singular}).

\section{Yangian Symmetry and Commuting Operators}
\setcounter{equation}{0}

The definite goal of this work is to construct algebraically the 
eigenstates of the Hubbard Hamiltonian (\ref{eqn:Hamiltonian}). 
As usual in the theory of integrable systems, we will not directly 
work with the Hamiltonian, but with an appropriately chosen generating 
function of a whole family of mutually commuting operators. For the
finite periodic system this generating function is the logarithm 
of the transfer matrix $\tau_{mn}(\lambda)$. 
Commuting operators are obtained as the coefficients of its expansion  
around $\lambda=0$ (\ref{eqn:expandtau}).
Equations (\ref{eqn:expandtildeT}) and (\ref{eqn:tildeL}) show that 
this expansion does not exist for the renormalized monodromy matrix
$\tilde{{\cal T}}(\lambda)$. There is substitute, however, which is 
intimately connected with the existence of an additional Y(su(2)) quantum 
group symmetry of the Hubbard model in the thermodynamic 
limit~\cite{MuGo97}. We shall describe it below. 

The commutation relations between the elements of the submatrix 
$A(\lambda)$ decouple from the rest of the algebra.
\begin{equation} 
     r(\lambda, \mu) \, (A (\lambda) \otimes A (\mu)) =
        (A(\mu) \otimes A (\lambda)) \, r(\lambda, \mu) ,
\label{eqn:commutA}
\end{equation}
where
\begin{equation}
     r(\lambda,\mu) = \frac{\rho_{3}\rho_{4}-\rho_{2}^{2}+\rho_{9}\rho_{10}
{\cal P}}{\rho_{4}
(\rho_{3}-\rho_{1})},
\label{eqn:matR}
\end{equation}
and ${\cal P}$ is a $4\times 4$ permutation matrix 
(${\cal P}x\otimes y=y\otimes x$).
If we introduce the reparametrization
\begin{eqnarray}
v(\lambda)&=&-2{\rm i} \, \cot 2 \lambda\cosh 2h(\lambda) \nonumber \\
&=&
-2\sin k(\lambda)+\frac{{\rm i}U}{2}=-2\sin p(\lambda)-\frac{{\rm i}U}{2},
\label{eqn:defv}
\end{eqnarray}
the $R$-matrix $r(\lambda,\mu)$ turns into the rational $R$-matrix 
of the isotropic Heisenberg spin chain,
\begin{equation}
r(\lambda,\mu)=\frac{{\rm i}U+(v(\lambda)-v(\mu)){\cal P}}{
{\rm i}U+v(\lambda)-v(\mu)}.
\end{equation}
Let us assume that $A(\lambda)$ allows for the following asymptotic 
expansion in 
terms of $v(\lambda)$, 
\begin{equation}
A(\lambda)=I_{2}+{\rm i}U \sum_{n=0}^{\infty}
\frac{1}{v(\lambda)^{n+1}}
\left( \sum_{a=1}^{3}
Q_{n}^{a}\tilde{\sigma}^{a}+Q_{n}^{0}I_{2}
\right),
\label{eqn:expandA}
\end{equation}
where $\tilde{\sigma}^{a}=G^{ab}\sigma^{b}$ (cf. (\ref{eqn:SigmaG})).
Then it follows from general considerations~\cite{Hal94,MuWa96b,BGHP}
that the first six operators $Q_{0}^{a},Q_{1}^{a}$ generate a 
representation of the Y(su(2)) Yangian quantum group.

There is the following alternative description of the Yangian 
Y(su(2))~\cite{Dri86}.
The Yangian Y(su(2)) is a Hopf algebra which is spanned by six 
generators $Q_{n}^{a} \ (n=0,1, \ a=x,y,z)$, satisfying the 
following relations, 
\begin{eqnarray}
& & \left[ Q_{0}^{a}, Q_{0}^{b} \right]  =  f^{abc}Q_{0}^{c}, 
\label{eqn:Y1} \\
& & \left[ Q_{0}^{a}, Q_{1}^{b} \right]  =  f^{abc}Q_{1}^{c}, 
\label{eqn:Y2} \\
& & \left[ \left[ Q_{1}^{a},Q_{1}^{b} \right], 
\left[ Q_{0}^{c},Q_{1}^{d} \right] \right]+
\left[ \left[ Q_{1}^{c},Q_{1}^{d} \right], 
\left[ Q_{0}^{a},Q_{1}^{b} \right] \right] \nonumber \\
& & \makebox[2em]{}
=\kappa^{2}(A^{abkefg}f^{cdk}+A^{cdkefg}f^{abk})
\{Q_{0}^{e}, Q_{0}^{f}, Q_{1}^{g} \}.
\label{eqn:Y3}
\end{eqnarray}
Here $\kappa$ is a nonzero constant, 
$f^{abc}=i\varepsilon^{abc}$ is the antisymmetric tensor of 
structure constants of su(2), and 
$A^{abcdef}=f^{adk}f^{bel}f^{cfm}f^{klm}$.
The bracket $\{\; \; \}$ in (\ref{eqn:Y3}) denotes the 
symmetrized product 
\begin{equation}
\{ x_{1}, x_{2}, x_{3}\} =\frac{1}{3!} \sum_{\sigma\in S_{3}}
x_{\sigma 1}x_{\sigma 2}x_{\sigma 3}.
\end{equation}
Being a Hopf algebra Y(su(2)) carries an outer structure 
(co-multiplication, antipode, co-unit), which is described in 
ref.~\cite{Dri86} and which assures that Y(su(2)) has a rich 
representation theory~\cite{ChaPre,ChaPreBOOK}.

A careful consideration of the limit $v(\lambda)\rightarrow\infty$ 
shows that $A(\lambda)$ is indeed of asymptotic form (\ref{eqn:expandA}).
There are several possibilities to carry out this limit as a 
function of $\lambda$. We found, however, that only one of these 
yields finite results for $Q_{0}^{a}$ and $Q_{1}^{a}$. 
We have to take $\Im (\lambda) \rightarrow \infty$ 
and have to choose the proper branch of 
solution of eq. (\ref{eqn:U4}), which determines $h$ as a function of
$\lambda$. 
(\ref{eqn:U4}) implies that
\begin{equation}
{\rm e}^{-2h(\lambda)}=-\frac{U}{4}\sin 2\lambda \pm 
\sqrt{1+\left( \frac{U}{4}\sin 2\lambda \right)^{2} }.
\label{eqn:e2h}
\end{equation}
in order to achieve convergence of the matrix elements 
$\tilde{\cal T}_{\alpha\beta} \; (\alpha,\beta=2,3)$
we have to choose the lower sign here, then 
${\rm e}^{-2h(\lambda)}$ is approximately 
equal to $-\frac{U}{2}\sin 2\lambda$
for large positive values of $\frac{U}{4}\sin 2\lambda$, 
and we obtain 
\begin{eqnarray}
&&{\rm e}^{2h(\lambda)}=
\frac{4{\rm i}}{U} {\rm e}^{2{\rm i}\lambda}
+O({\rm e}^{6{\rm i}\lambda}), \makebox[1em]{}
{\rm e}^{{\rm i}k(\lambda)}=
\frac{-4}{U} \{ {\rm e}^{2{\rm i}\lambda}
+2{\rm e}^{4{\rm i}\lambda}+
O({\rm e}^{6{\rm i}\lambda}) \}, \nonumber \\
&&{\rm e}^{-{\rm i}p(\lambda)}=
\frac{4}{U} \{ {\rm e}^{2{\rm i}\lambda}
-2{\rm e}^{4{\rm i}\lambda}+O({\rm e}^{6{\rm i}\lambda}) \}, 
\makebox[1em]{}
\frac{1}{v(\lambda)}=\frac{-4{\rm i}}{U}
\{ {\rm e}^{2{\rm i}\lambda}+
O({\rm e}^{6{\rm i}\lambda}) \}.
\label{eqn:expandfunctions}
\end{eqnarray}
The leading terms in the series (\ref{eqn:expandtildeT})
are of order ${\rm e}^{2{\rm i}\lambda}$, 
${\rm e}^{4{\rm i}\lambda}, \cdots$. 
Thus, 
from the first two sums in (\ref{eqn:expandtildeT}),
we get the expansion of the matrix $A(\lambda)$ up to order 
${{\rm e}}^{4{{\rm i}}\lambda}$, and
the last equation in (\ref{eqn:expandfunctions})
yields the required expansion in $(v(\lambda))^{-1}$ up to 
second order.

The final result for the representation of Yangian generators is 
\begin{eqnarray}
Q_{0}^{a}&=& \frac{1}{2}\sum_{j}\sigma_{\alpha\beta}^{a}
c^{\dagger}_{j,\alpha}c_{j,\beta}, 
\label{eqn:Q0}
\\
Q_{1}^{a}&=& -\frac{{\rm i}}{2}\sum_{j}\sigma_{\alpha\beta}^{a}
c_{j,\alpha}^{\dagger}(c_{j+1,\beta}-c_{j-1,\beta})
-\frac{{\rm i}U}{4}\sum_{i,j}
{\rm sgn}(j-i)\sigma_{\alpha\beta}^{a}
c_{i,\alpha}^{\dagger}c_{j,\gamma}^{\dagger}
c_{i,\gamma}c_{j,\beta}.
\label{eqn:Q1}
\end{eqnarray}
The factor ${\rm i}U$ occurring in (\ref{eqn:Q1}) can be 
identified with the constant $\kappa$ in (\ref{eqn:Y3}). 
Note that $Q_{0}^{a}=S^{a}$ is just the operator of the 
$a$-component of the total spin (cf. (\ref{eqn:spin})). 
The Yangian representation (\ref{eqn:Q0}) and 
(\ref{eqn:Q1}) was first obtained by 
Uglov and Korepin~\cite{UgKo94,MuGo97}. It can be embedded 
into a larger family of Yangian representations connected with 
long-range-hopping extensions of the Hamiltonian 
(\ref{eqn:Hamiltonian})~\cite{GoIn96}. 
Uglov and Korepin showed that $Q_{0}^{a}$ and $Q_{1}^{a}$ commute 
with the Hamiltonian on the infinite line.

Since the quantum determinant
\begin{equation}
{\rm Det}_{q}A(\lambda)=
A_{11}(\lambda)A_{22}(\check{\lambda})-
A_{12}(\lambda)A_{21}(\check{\lambda}),
\end{equation}
where $v(\check{\lambda})=v(\lambda)-{\rm i}U$,
is in the center of the Yangian 
\begin{equation}
[{\rm Det}_{q}A(\lambda), A(\mu)]=0, 
\label{eqn:DetAA}
\end{equation}
and thus provides a generating
function of mutually commuting operators,
\begin{equation}
[{\rm Det}_{q}A(\lambda), {\rm Det}_{q}A(\mu)]=0,
\end{equation}
it is a natural candidate to generate the Hamiltonian, too. 
Performing again the asymptotic expansion in
terms of $v(\lambda)^{-1}$, 
\begin{equation}
{\rm Det}_{q}A(\lambda)= 1+{\rm i}U\sum_{n=0}^{\infty}
\frac{J_{n}}{v(\lambda)^{n+1}}, 
\label{eqn:expdetA}
\end{equation}
we obtain 
$J_{0}=0$, $J_{1}={\rm i}\hat{H}$, 
i.e.\ the Hamiltonian is indeed among the commuting operators 
generated by ${\rm Det}_{q}A(\lambda)$.
All the conserved operators are Yangian invariant by construction. 
We discuss their relation to the formerly known conserved 
quantities~\cite{Sha88,OlWa88,Gro,GraMa} in section 5.2 below.

In closing this section we shall add a comment. 
The Hubbard Hamiltonian on the infinite interval is invariant 
(up to a constant) under 
the transformation (\ref{eqn:transformation}).
The Yangian generators $Q_{0}^{a}$ and $Q_{1}^{a}$, 
however, are transformed into a pair of generators
$Q_{0}^{a\prime}$ and $Q_{1}^{a\prime}$ of a second, independent 
representation of Y(su(2))~\cite{UgKo94}. These two representations 
mutually commute. Therefore they can be combined to a 
direct sum Y(su(2))$\oplus$Y(su(2)). The reason why we get only 
one of these representations from our QISM approach is that, 
in order to perform the passage to the infinite interval, we 
refer to the zero density vacuum $|0\rangle$. This vacuum has lower 
symmetry than the Hamiltonian. It is invariant under the su(2) Lie 
algebra of rotations, but does not respect the $\eta$-pairing 
su(2) symmetry of the Hamiltonian. 
A fully su(2)$\oplus$su(2) invariant vacuum would be the singlet
ground state at half filling~\cite{EsKo94}.
It seems to be yet a formidable task to formulate the 
QISM with respect to this state.

\section{Conserved Quantities and Eigenvectors}
\setcounter{equation}{0}

\subsection{Eigenvectors of the Hamiltonian}

In the present section we will try to understand
the meaning of the operators contained in $B(\lambda),C(\lambda),
D(\lambda)$. To begin with, let us have a look at the commutator of 
$\tilde{{\cal T}}(\lambda)$ with the particle number operator 
$\hat{N}$, which is the extension of $\hat{N}_{nm}$ (\ref{eqn:N})
to the infinite interval. Note that 
$[V(\lambda),\tilde{\Sigma}_{\eta}^{z}]=0$, and thus by 
(\ref{eqn:Teta})
\begin{equation}
  [\hat{N},\tilde{{\cal T}}(\lambda)]
=2[\tilde{{\cal T}}(\lambda),\tilde{\Sigma}_{\eta}^{z}].
\end{equation}
This is a set of 16 equations. Writing out the equations explicitly, 
we find that $D_{11}(\lambda)$, $D_{22}(\lambda)$ and 
$A(\lambda)$ conserve the number of particles. The 
operators $B_{a1}(\lambda)$ and $C_{2a}(\lambda)$ increase the number 
of particles by one, whereas $B_{a2}(\lambda)$ and $C_{1a}(\lambda)$ 
reduce the number of particles by one. $D_{21}(\lambda)$ adds two 
particles to the system, whereas $D_{12}(\lambda)$ removes 
two particles. Hence, $B_{a1}(\lambda)$,$C_{2a}(\lambda)$ and 
$D_{21}(\lambda)$ are candidates for creation operators of eigenstates 
of the Hamiltonian. The action of these operators on the vacuum follows 
from (\ref{eqn:expandtildeT}),
\begin{eqnarray}
\label{eqn:B11v}
&&B_{11} (\lambda)
|0\rangle = - \frac{{\rm i}{\rm e}^{h(\lambda)}}{\cos \lambda} 
\sum_m {\rm e}^{- {\rm i}mp(\lambda)}
c_{m \downarrow}^{\dagger} |0\rangle , \\
\label{eqn:B12v}
&&B_{21} (\lambda) |0\rangle  =  
\frac{{\rm e}^{h(\lambda)}}{\cos\lambda } \sum_m {\rm e}^{- {\rm i}mp(\lambda)}
c_{m \uparrow}^{\dagger} |0\rangle , \\
\label{eqn:C21v}
&&C_{21} (\lambda) |0\rangle  =  - \frac{{\rm e}^{-h(\lambda)}}{\cos\lambda} 
\sum_m {\rm e}^{- {\rm i}mk(\lambda)}
c_{m \uparrow}^{\dagger} |0\rangle , \\
\label{eqn:C22v}
&&C_{22} (\lambda) |0\rangle  = - 
\frac{{\rm i}{\rm e}^{-h(\lambda)}}{\cos\lambda} 
\sum_m {\rm e}^{- {\rm i}mk(\lambda)}
c_{m \downarrow}^{\dagger} |0\rangle , \\
\label{eqn:D21v}
&&D_{21} (\lambda) |0\rangle =  \frac{{\rm i}}{\cos^2 \lambda} 
\sum_{m,n} c_{m \uparrow}^{\dagger} c_{n \downarrow}^{\dagger}
\left\{
\theta(m\geq n){\rm e}^{- {\rm i}(mk(\lambda) + np(\lambda))}
+
\theta(m< n){\rm e}^{- {\rm i}(nk(\lambda) + mp(\lambda))}
\right\} |0\rangle .
\end{eqnarray}
These states are the most elementary eigenstates of the Hamiltonian
as can be verified by direct calculation. 
Note that they are bounded in different, disconnected parts of the 
spectral parameter space. 
This is because of the constraint (\ref{eqn:U4}), which turns into 
\begin{equation}
\sin k(\lambda) -\sin p(\lambda)=\frac{{\rm i}U}{2}  
\label{eqn:sinksinp}
\end{equation}
when expressed in terms of $k(\lambda)$, $p(\lambda)$ (\ref{eqn:defkp}).
If 
$k(\lambda)$ is real, $p(\lambda)$ cannot be real for non-zero $U$
and vice versa. $D_{21} (\lambda)$ creates a bound state, if
$\Im (k) = - \Im (p) < 0$.
This condition is compatible with the constraint (\ref{eqn:sinksinp}) on 
$k(\lambda)$ and $p(\lambda)$. 
Let $U\ne 0$. Then $\kappa := \Im (p)
= - \Im (k)$ implies $q := \Re (k) = \Re (p) \, {\rm mod} \, 2\pi$, and
\begin{equation} 
\label{eqn:constr}
\sinh \kappa = - \frac{U}{4\cos q} .
\end{equation}
Hence, if $U < 0$, $D_{21} (\lambda)$ creates a bound state for $|q| < \p/2$,
and if $U > 0$, $D_{21} (\lambda)$ creates a bound state for $\p/2 < |q| < \p$.

The commutators of the various operators contained in 
$\tilde{{\cal T}}(\lambda)$ with ${\rm Det}_{q}(A(\mu))$ are summarized in
\ref{appendix:list}. Let us insert the asymptotic expansion
(\ref{eqn:expdetA}) into (\ref{eqn:detB1})-(\ref{eqn:detD2}). 
Then there occur only 
two different ratios of Boltzmann weights in these equations, which 
may be expanded by use of 
(\ref{eqn:expandfunctions}) as
\begin{eqnarray}
\label{eqn:exp1}
- \frac{{\rm i}\rho_1 (\lambda, \mu)}{\rho_9 (\lambda, \mu)}
 & = & 1 + \frac{{\rm i}U}{2v(\mu)}
+ \left( \frac{{\rm i}U}{8} - {\rm i}e^{- {\rm i}k(\lambda)} \right)
\frac{{\rm i}U}{v(\mu)^2} + O \left( \frac{1}{v(\mu)^3} \right) 
, \\
\label{eqn:exp2}
\frac{{\rm i}\rho_{10} (\lambda, \mu)}{\rho_4 (\lambda, \mu)} & = & 1 
+ \frac{{\rm i}U}{2v(\mu)}
+ \left( \frac{{\rm i}U}{8} + {\rm i}e^{{\rm i}p(\lambda)} \right)
\frac{{\rm i}U}{v(\mu)^2} + O \left( \frac{1}{v(\mu)^3} \right) .
\end{eqnarray}
Comparing terms of second order in $v(\mu)^{-1}$, 
we obtain the following commutators, 
\begin{eqnarray}
\label{eqn:HB1}
[\hat H, B_{a1} \lambda)] & = & - (2\cos p(\lambda) + U/2) B_{a1} (\lambda)
, \\
\label{eqn:HB2}
[\hat H, B_{a2} 
(\lambda)] & = & (2\cos k(\lambda) + U/2) B_{a2} (\lambda)
, \\
\label{eqn:HC1}
[\hat H, C_{1a} \lambda)] & = & (2\cos p(\lambda) + U/2) C_{1a} (\lambda)
, \\
\label{eqn:HC2}
[\hat H, C_{2a}(\lambda)] & = & - (2\cos k(\lambda) + U/2) C_{2a} (\lambda)
 , \\
\label{eqn:HD1}
[\hat H, D_{12} (\lambda)] & = & 
2({\rm e}^{{\rm i}p(\lambda)} + {\rm e}^{- {\rm i}k(\lambda)})
D_{12} (\la) , \\
\label{eqn:HD2}
[\hat H, D_{21} (\lambda)] & = & 
- 2({\rm e}^{{\rm i}p(\lambda)} + {\rm e}^{- {\rm i}k(\lambda)})
D_{21} (\lambda) .
\end{eqnarray}
The above results justify the interpretation of 
$B_{a1}(\lambda)$,$C_{2a}(\lambda)$ and $D_{21}(\lambda)$ as creation 
operators. 
$B_{a1}(\lambda)$ and $C_{2a}(\lambda)$ create single particle 
excitations, whereas $D_{21}(\lambda)$ creates a bound state of two 
particles. Let us investigate several examples. 
Since $\hat{H}|0\rangle=0$, (\ref{eqn:HC2}) implies, for instance,
\begin{equation}
\hat H C_{2a} (\lambda) |0\rangle  = - (2\cos k(\lambda) + U/2) \,
C_{2a} (\lambda) |0\rangle ,
\end{equation}
or more generally
\begin{equation}
\hat H C_{2a_1} (\lambda_1) \dots C_{2a_n} (\lambda_n) |0\rangle =
- \sum_{j=1}^n (2\cos k(\lambda_j) + U/2) \,
C_{2a_1} (\lambda_1) \dots C_{2a_n} (\lambda_n) |0\rangle .
\end{equation}
A similar result holds for states, where the operators $B_{1a} (\lambda)$
or mixed products of operators $B_{a1} (\lambda)$ and $C_{2a} (\lambda)$ are
applied to the vacuum. We have to remember 
here that $B_{a1}(\lambda)$ and $C_{2a}(\mu)$ create bounded states, 
only if $p(\lambda)$ and $k(\mu)$ are real. 
The restrictions on $k(\lambda)$ and 
$p(\lambda)$ occurring in (\ref{eqn:HD2}) have
been, $k = q - {\rm i}\kappa$, $p = q + {\rm i}\kappa$, 
where $q$ is real and $\kappa > 0$.
Taking these restrictions into account we obtain, for instance,
\begin{equation}
\hat H D_{21} (\lambda) |0\rangle = - (4\cos q \cosh \kappa + U) \,
D_{21} (\lambda) |0\rangle .
\end{equation}
The constraint (\ref{eqn:sinksinp}) implies
\begin{equation}
4\cos q \cosh\kappa = \pm \sqrt{16 \cos^2 q + U^2} ,
\end{equation}
where the plus sign has to be taken for $U < 0$, $|q| < \frac{\pi}{2}$, and
the minus sign is relevant, if $U > 0$ and $\frac{\pi}{2} < |q| < \pi$. This
is, of course, in accordance with intuition.

\subsection{Higher Conserved Quantities}

Within the present formalism it is of course more natural to consider 
the quantum determinant ${\rm Det}_{q}(A(\lambda))$ rather than the 
Hamiltonian.
It can be seen from the commutation relations in \ref{appendix:list} that 
arbitrary products of operators $B_{a1}(\lambda)$, $C_{2a}(\lambda)$ 
and $D_{21}(\lambda)$ create eigenstates of the quantum determinant 
${\rm Det}_{q}A(\lambda)$. For example, (\ref{eqn:detC2}) and 
${\rm Det}_{q}A(\mu)|0\rangle=|0\rangle$ yield
\begin{equation}
{\rm Det}_{q}(A(\mu))C_{2a}(\lambda)|0\rangle =
      - \frac{\r_1 (\lambda, \mu) \r_1 (\lambda, \check{\mu})}
             {\r_9 (\lambda, \mu) \r_9 (\lambda, \check{\mu})}
     \frac{v(\lambda)-v(\mu) + {\rm i}U}{v(\lambda)-v(\mu)} \, 
C_{2a} (\lambda) |0\rangle,
\label{eqn:DetC2v}
\end{equation}
where $\check{\mu}$ is defined by $v(\check{\mu})=v(\mu)-{\rm i}U$.
The eigenvalue of multi-particle states is a product of eigenvalues of 
single particle states. For this reason 
$\ln {\rm Det}_{q}(A(\mu))$ is a generating function of conserved 
operators, which is possessing an additive spectrum on multi-particle 
states. 

Commuting conserved operators for the Hubbard model have been 
constructed by several authors~\cite{Sha88,OlWa88,Gro,GraMa} 
either by use of ad hoc
methods or by using the expansion (\ref{eqn:expandtau}). We present the 
first few known of them in \ref{appendix:cons}. Here we ask for 
the relation of these known conserved operators to the ones 
generated by $\ln {\rm Det}_{q}(A(\mu))$. To this end let us 
compare the action on one-particle states. 
Equation (\ref{eqn:DetC2v}) implies
\begin{equation}
\ln {\rm Det}_{q}(A(\mu))\sum_{j}{\rm e}^{-{\rm i}kj}
c^{\dagger}_{j\sigma}|0\rangle =
\Upsilon(\mu,k) \sum_{j}{\rm e}^{-{\rm i}kj}
c^{\dagger}_{j\sigma}|0\rangle,
\label{eqn:expandlndet}
\end{equation}
where
\begin{equation}
\Upsilon(\mu,k)=\ln\left[ 
\frac{\displaystyle\cos\frac{k+p(\mu)}{2}}{\displaystyle\sin\frac{k-k(\mu)}{2}}
\ \frac{\displaystyle
\cos\frac{k+p(\check{\mu})}{2}}{\displaystyle\sin\frac{k-k(\check{\mu})}{2}}
\ \frac{\sin k-\sin k(\mu)-{\rm i}U/2}{\sin k-\sin k(\mu)}
\right].
\end{equation}

Comparing the expansion of $\Upsilon(\mu,k)$ in terms of 
$v(\lambda)^{-1}$ with the eigenvalues of the first few explicitly known 
higher conserved quantities $H_{1}(=\hat{H})$ ,
$H_{2}$, $H_{3}$, $H_{4}$ (for details see 
\ref{appendix:cons}), we are led to the following conjecture,
\begin{eqnarray}
&&\ln {\rm Det}_{q}(A(\mu))\nonumber \\
&&\makebox[2em]{}
=\frac{{\rm i}U}{v(\mu)^2}{\rm i}H_{1}
+\frac{{\rm i}U}{v(\mu)^3}({\rm i}H_{2}-U H_{1})
\nonumber \\
&&\makebox[2em]{}
+\frac{{\rm i}U}{v(\mu)^4}\left[ {\rm i}H_{3}-\frac{3U}{2}H_{2}+
\left(\frac{-5{\rm i}}{4}U^{2}+3{\rm i}\right)H_{1}\right] \nonumber \\
&&\makebox[2em]{}
+\frac{{\rm i}U}{v(\mu)^5}\left[ {\rm i}H_{4}-2U H_{3}+
\left(-\frac{11{\rm i}}{4}U^{2}+4{\rm i}\right)H_{2}+
\left(\frac{3U^{3}}{2}-6U\right)H_{1}\right] \nonumber \\
&&\makebox[2em]{}+O\left(\frac{1}{v(\mu)^{6}}\right).
\label{eqn:lndetA}
\end{eqnarray}
The coefficients of the asymptotic expansion of 
$\ln{\rm Det}_{q}(A(\mu))$ are linear combinations of the formerly known 
conserved quantities. Particularly, they do not give a complete set of 
conserved operators of the free fermion model in the limit 
$U\rightarrow 0$ (cf.\ \ref{appendix:cons}). It remains therefore an open 
question, if $\ln{\rm Det}_{q}(A(\lambda))$ generates a complete set of 
commuting operators for the Hubbard model or not. 
Another fact which can be concluded from (\ref{eqn:lndetA}) is the 
Yangian invariance of the operators $H_{s}$ (cf.\ \cite{ShUjWa97}).

At this point a remark on the construction of Dunkl operators may 
be in order. Dunkl operators are commuting difference (differential)
operators, which turned out to be useful in the investigation of exactly 
solvable long-range interacting systems~\cite{Dunkl,BGHP}. More precisely, 
a set of difference operators $\{ d_{j}\}$ is a set of Dunkl operators, 
if there exists a representation $\{ K_{ij}\}$ of the 
symmetric group, such that the operators $d_{j},K_{ij}$ form a
representation of the degenerate affine Hecke algebra. 
Given a representation of the degenerate affine Hecke algebra, it is 
possible to construct a corresponding representation of the Y(su(2)) 
Yangian out of it~\cite{Dunkl,BGHP}. The strategy developed in 
ref.~\cite{Dunkl,BGHP} 
was successfully applied to the fermionic nonlinear Schr\"odinger 
model~\cite{MuWa96b}. Yet it seems to be inappropriate for the 
Hubbard model for the following reason. 
Let $P_{0j}$ be a permutation operator acting on su($p$) spins. Then 
it follows from the defining relations of the degenerate affine 
Hecke algebra that the transfer matrix
\begin{equation}
\hat{T}_{0}(u)=\left(
1+\frac{{\rm i}c\ P_{01}}{u-d_{1}}
\right)
\cdots
\left(
1+\frac{{\rm i}c\ P_{0n}}{u-d_{n}}
\right)
\end{equation}
preserves the space of fermionic wave functions. It satisfies the Yang-Baxter 
relations with $R$-matrix $u+{\rm i}cP$ and thus generates a representation 
of Y(su($p$)). In case of the fermionic nonlinear Schr\"odinger model 
the quantum determinant ${\rm Det}_{q}(\hat{T}(u))$ agrees with 
the quantum determinant of a submatrix of the monodromy matrix obtained 
within the quantum inverse scattering approach~\cite{MuWa96b}. 
It has the eigenvalue
\begin{equation}
\prod_{j=1}^{n}\left(
1+\frac{{\rm i}c}{u-k_{j}} \right),
\end{equation}
By way of contrast the eigenvalue of the quantum determinant 
${\rm Det}_{q}(A(\mu))$ is 
\begin{equation}
\prod_{j=1}^{n}\exp\Upsilon(\mu,k_{j})=\prod_{j=1}^{n}\left[ 
\frac{\displaystyle\cos\frac{k_{j}+p(\mu)}{2}}{\displaystyle
\sin\frac{k_{j}-k(\mu)}{2}}
\ \frac{\displaystyle
\cos\frac{k_{j}+p(\check{\mu})}{2}}{\displaystyle
\sin\frac{k_{j}-k(\check{\mu})}{2}}
\ \frac{\sin k_{j}-\sin k(\mu)-{\rm i}U/2}{\sin k_{j}-\sin k(\mu)}
\right].
\end{equation}
Hence it seems that the method developed in ref.~\cite{MuWa96b} has to be 
modified, if we want to apply it to the Hubbard model.

\subsection{The elements of the monodromy matrix under Yangian transformations}
As we have shown above, the submatrix $A(\lambda)$ of the monodromy matrix
$\tilde{\cal T}(\lambda)$ generates a representation of Y(su(2)). 
Let us look 
for the commutators of the remaining elements of the monodromy matrix, which
can be arranged in submatrices $B(\lambda)$, $C(\lambda)$, $D(\lambda)$,
with the Yangian generators $Q_n^a$, $n = 0, 1$; $a = x, y, z$. 
Combining (\ref{eqn:expandA}) and (\ref{eqn:BA1})-(\ref{eqn:DA2})
we end up with
\begin{eqnarray}
\label{eqn:YB0}
[Q_{0}^{a}, B (\lambda)]& = & -\frac{1}{2}\tilde{\sigma}^{a} B (\lambda) , \\
 \label{eqn:YB1}
[Q_{1}^{a}, B (\lambda)] & = & \sin p(\lambda) \tilde{\sigma}^{a} B (\lambda)
+ \frac{U}{2} \varepsilon^{abc}
\tilde{\sigma}^{b} B (\lambda) Q_{0}^{c} , \\
\label{eqn:YC0}
[Q_{0}^{a}, C (\lambda)] & = & \frac{1}{2} C (\lambda) \tilde{\sigma}^{a} , \\
\label{eqn:YC1}
[Q_{1}^{a}, C (\lambda)] & = & - \sin k(\lambda)C(\lambda)\tilde{\sigma}^{a}
+\frac{U}{2} \varepsilon^{abc}
C (\lambda) \tilde{\sigma}^{b}Q_{0}^{c} , \\
\label{eqn:YD}
[Q_{0}^{a}, D (\lambda)] & = & [Q_{1}^{a}, D (\lambda)] \, = \, 0 .
\end{eqnarray}
In the next section we will see that these equations 
determine the behavior of the eigenstates of the
Hamiltonian under Yangian transformations. The
discussion of the irreducible representations on the subspace of a
fixed number of one-particle excitations created by $C_{2a} (\lambda)$
can be done in analogy with ref.~\cite{MuWa96b}.

\section{Construction of $n$-particle states}
\setcounter{equation}{0}

\subsection{Scattering States}

We have seen in the preceding section that the repeated action of 
operators $B_{1a}(\lambda)$, $C_{2a}(\lambda)$ on the vacuum produces 
$n$-particle eigenstates of the quantum determinant of $A(\lambda)$. 
For small enough $n$ the corresponding wave functions can be 
worked out by hand. They are of the form of Bethe wave functions 
and are easily understood as scattering states of $n$-particles. 
For scattering states there is a natural normalization. We 
have to require the amplitude of the incident wave to be unity. 
In previously studied cases~\cite{Fa80,Tha81,PuWuZh87} 
it turned out that such kind of normalization was 
obtainable by introducing the operator analog of the reflection 
coefficient of the corresponding classical problem. 
In other words, the creation operators were normalized 
by multiplying with the inverse of certain generators of conserved 
quantities. 

For the Hubbard model we propose the following two pairs of 
normalized creation operators, 
\begin{eqnarray}
R_{\alpha}(\lambda)^{\dagger}
&=&{\rm i}^{3-\alpha}{\rm e}^{h(\lambda)}\cos \lambda \ 
C_{2\alpha}(\lambda)D_{22}(\lambda)^{-1} \  (\alpha=1,2),
\label{eqn:defRd}\\
\hat{R}_{\alpha}(\lambda)^{\dagger}
&=&{\rm i}^{\alpha-1}{\rm e}^{-h(\lambda)}\cos \lambda \ 
B_{3-\alpha,1}(\lambda)D_{11}(\lambda)^{-1} \ (\alpha=1,2).
\label{eqn:defRhd}
\end{eqnarray}
In these formulae 
$\alpha=1$ corresponds to spin-up and $\alpha=2$ 
to spin-down, respectively.
The numerical prefactors have been obtained by demanding that 
$R_{\alpha}(\lambda)^{\dagger}$ and 
$\hat{R}_{\alpha}(\lambda)^{\dagger}$ 
generate normalized one-particle states above the vacuum,
\begin{equation}
R_{\alpha}(\lambda)^{\dagger}|0\rangle=\sum_{m}{\rm e}^{-{\rm i}mk(\lambda)}
c_{m,\alpha}^{\dagger}|0\rangle, \ \  \
\hat{R}_{\alpha}(\lambda)^{\dagger}
|0\rangle=\sum_{m}{\rm e}^{-{\rm i}mp(\lambda)}
c_{m,\alpha}^{\dagger}|0\rangle.
\label{eqn:Roneparticle}
\end{equation}
Hereafter we assume that $\lambda$ is chosen in such a way that 
$R_{\alpha}(\lambda)^{\dagger}$ and $\hat{R}_{\alpha}(\lambda)^{\dagger}$ 
create physical states. This means for $R_{\alpha}(\lambda)^{\dagger}$ 
that $k(\lambda)$ has to be real and for $\hat{R}_{\alpha}
(\lambda)^{\dagger}$ that $p(\lambda)$ has to be real.

It is not difficult to see that eqs.~(\ref{eqn:Adagger})-(\ref{eqn:Ddagger}), 
which determine the behavior of the elements of the monodromy matrix 
under hermitian conjugation, remain valid in the thermodynamic limit.  
Hence, they can be used to obtain the conjugated annihilation 
operators corresponding to $R_{\alpha}(\lambda)^{\dagger}$ and 
$\hat{R}_{\alpha}(\lambda)^{\dagger}$,
\begin{eqnarray}
R_{\alpha}(\lambda)
&=&{\rm i}^{2-\alpha}{\rm e}^{h(\lambda')}\sin \lambda' \ 
D_{11}(\lambda')^{-1}C_{1,3-\alpha}(\lambda') \  (\alpha=1,2),
\label{eqn:defR}\\
\hat{R}_{\alpha}(\lambda)
&=&{\rm i}^{\alpha-2}{\rm e}^{-h(\lambda')}\sin \lambda' \ 
D_{22}(\lambda')^{-1}B_{\alpha 2}(\lambda') \ (\alpha=1,2),
\label{eqn:defRh}
\end{eqnarray}
where $\lambda'=\pi/2-\lambda^{*}$.
The commutation relations between the normalized operators 
are easily calculated by use of the formulae presented in 
\ref{appendix:list}. Provided $\lambda\neq\mu \ ({\rm mod} \ 2\pi)$
the results are
\begin{eqnarray}
R_{\alpha}(\lambda)^{\dagger}R_{\beta}(\mu)^{\dagger}
&=&-r(\lambda,\mu)_{\gamma\delta,\alpha\beta}R_{\gamma}(\mu)^{\dagger}
R_{\delta}
(\lambda)^{\dagger},\label{eqn:RdRd}
\\
R_{\alpha}(\lambda)R_{\beta}(\mu)^{\dagger}
&=&-r(\mu,\lambda)_{\gamma\alpha,\delta\beta}R_{\gamma}(\mu)^{\dagger}
R_{\delta}
(\lambda),\label{eqn:RRd}
\\
\hat{R}_{\alpha}(\lambda)^{\dagger}\hat{R}_{\beta}(\mu)^{\dagger}
&=&-r(\mu,\lambda)_{\gamma\delta,\alpha\beta}
\hat{R}_{\gamma}(\mu)^{\dagger}\hat{R}_{\delta}(\lambda)^{\dagger},
\label{eqn:RhdRhd}\\
\hat{R}_{\alpha}(\lambda)\hat{R}_{\beta}(\mu)^{\dagger}
&=&-r(\lambda,\mu)_{\gamma\alpha,\delta\beta}
\hat{R}_{\gamma}(\mu)^{\dagger}\hat{R}_{\delta}(\lambda),
\label{eqn:RhRhd}\\
R_{\alpha}(\lambda)^{\dagger}\hat{R}_{\beta}(\mu)^{\dagger}
&=&-\hat{R}_{\beta}(\mu)^{\dagger}R_{\alpha}(\lambda)^{\dagger}, 
\label{eqn:RdRhd}\\
R_{\alpha}(\lambda)\hat{R}_{\beta}(\mu)^{\dagger}
&=&-\hat{R}_{\beta}(\mu)^{\dagger}R_{\alpha}(\lambda).
\label{eqn:RRhd}
\end{eqnarray}

The operators $R_{\alpha}(\lambda)$, $R_{\alpha}(\lambda)^{\dagger}$
and 
$\hat{R}_{\alpha}(\lambda)$, $\hat{R}_{\alpha}(\lambda)^{\dagger}$
form a representation of the graded Zamolodchikov-Faddeev algebra
with $S$-matrix $r(\lambda,\mu)$. These representations may be 
identified as representations of left and right Zamolodchikov-Faddeev 
algebra, respectively~\cite{ZaZa79,Fa80,Kha,SmiBOOK,GoRuSiBOOK}. 
The grading is such that all operators are odd. In physical 
terms we may say that $R_{\alpha}(\lambda)^{\dagger}$
and $\hat{R}_{\alpha}(\lambda)^{\dagger}$ are creation operators 
of fermionic quasi-particles. 
Both quasi-particles are charge density waves, because 
$R_{\alpha}(\lambda)^{\dagger}$ and $\hat{R}_{\alpha}(\lambda)^{\dagger}$
add a particle to the system. Note that we cannot have spin 
density waves over the zero density vacuum.

It is possible to calculate the $S$-matrix for two-body scattering processes 
within the coordinate Bethe ansatz from two-body phase 
shifts~\cite{Ko79,AnLo80}. This approach was applied to the Hubbard model 
at half-filling by E{\ss}ler and Korepin~\cite{EsKo94}. The presence of a
finite density background of particles influences the scattering. 
Thus the $S$-matrix of E{\ss}ler and Korepin for holon-holon 
scattering differs from ours by a dressing factor. 

Let us make our above statements about the creation of normalized 
scattering states more precise. 
We shall present the two-particle states generated by 
$R_{\alpha}(\lambda)^{\dagger}$ 
or $\hat{R}_{\alpha}(\lambda)^{\dagger}$ as derived from 
(\ref{eqn:expandtildeT}) and the 
commutation relations between the elements of the monodromy matrix.
\begin{eqnarray}
&&R_{1}(\lambda)^{\dagger}R_{1}(\mu)^{\dagger}|0\rangle=
\sum_{n,m}c_{n\uparrow}^{\dagger}c_{m\uparrow}^{\dagger}
{\rm e}^{-{\rm i}nk(\lambda)}{\rm e}^{-{\rm i}mk(\mu)}|0\rangle, 
\label{eqn:R1R1} \\
&&R_{2}(\lambda)^{\dagger}R_{2}(\mu)^{\dagger}|0\rangle=
\sum_{n,m}c_{n\downarrow}^{\dagger}c_{m\downarrow}^{\dagger}
{\rm e}^{-{\rm i}nk(\lambda)}{\rm e}^{-{\rm i}mk(\mu)}|0\rangle, 
\label{eqn:R2R2}\\
&&R_{1}(\lambda)^{\dagger}R_{2}(\mu)^{\dagger}|0\rangle=
\sum_{n,m}c_{n\uparrow}^{\dagger}c_{m\downarrow}^{\dagger}
\left[ \theta(n\geq m){\rm e}^{-{\rm i}nk(\lambda)}{\rm e}^{-{\rm i}mk(\mu)}
\frac{v(\lambda)-v(\mu)}{v(\lambda)-v(\mu)+{\rm i}U} \right.
\nonumber \\
&&\makebox[2em]{}
\left. +\theta(n< m){\rm e}^{-{\rm i}nk(\lambda)}{\rm e}^{-{\rm i}mk(\mu)}
+\theta(n< m){\rm e}^{-{\rm i}mk(\lambda)}{\rm e}^{-{\rm i}nk(\mu)}
\frac{-{\rm i}U}{v(\lambda)-v(\mu)+{\rm i}U} \right]|0\rangle, \makebox[2em]{}
\label{eqn:R1R2} \\
&&R_{2}(\lambda)^{\dagger}R_{1}(\mu)^{\dagger}|0\rangle=
\sum_{n,m}c_{n\downarrow}^{\dagger}c_{m\uparrow}^{\dagger}
\left[ \theta(n\geq m){\rm e}^{-{\rm i}nk(\lambda)}{\rm e}^{-{\rm i}mk(\mu)}
\frac{v(\lambda)-v(\mu)}{v(\lambda)-v(\mu)+{\rm i}U} \right.
\nonumber \\
&&\makebox[2em]{}
\left. +\theta(n< m){\rm e}^{-{\rm i}nk(\lambda)}{\rm e}^{-{\rm i}mk(\mu)}
+\theta(n< m){\rm e}^{-{\rm i}mk(\lambda)}{\rm e}^{-{\rm i}nk(\mu)}
\frac{-{\rm i}U}{v(\lambda)-v(\mu)+{\rm i}U} \right]|0\rangle, \makebox[2em]{}
\label{eqn:R2R1}\\
&&\hat{R}_{1}(\lambda)^{\dagger}\hat{R}_{1}(\mu)^{\dagger}|0\rangle=
\sum_{n,m}c_{n\uparrow}^{\dagger}c_{m\uparrow}^{\dagger}
{\rm e}^{-{\rm i}np(\lambda)}{\rm e}^{-{\rm i}mp(\mu)}|0\rangle, 
\label{eqn:Rh1Rh1}\\
&&\hat{R}_{2}(\lambda)^{\dagger}\hat{R}_{2}(\mu)^{\dagger}|0\rangle=
\sum_{n,m}c_{n\downarrow}^{\dagger}c_{m\downarrow}^{\dagger}
{\rm e}^{-{\rm i}np(\lambda)}{\rm e}^{-{\rm i}mp(\mu)}|0\rangle, 
\label{eqn:Rh2Rh2} \\
&&\hat{R}_{1}(\lambda)^{\dagger}\hat{R}_{2}(\mu)^{\dagger}|0\rangle=
\sum_{n,m}c_{n\uparrow}^{\dagger}c_{m\downarrow}^{\dagger}
\left[ \theta(n\leq m){\rm e}^{-{\rm i}np(\lambda)}{\rm e}^{-{\rm i}mp(\mu)}
\frac{v(\lambda)-v(\mu)}{v(\lambda)-v(\mu)-{\rm i}U} \right.
\nonumber \\
&&\makebox[2em]{}
\left. +\theta(n> m){\rm e}^{-{\rm i}np(\lambda)}{\rm e}^{-{\rm i}mp(\mu)}
+\theta(n> m){\rm e}^{-{\rm i}mp(\lambda)}{\rm e}^{-{\rm i}np(\mu)}
\frac{{\rm i}U}{v(\lambda)-v(\mu)-{\rm i}U} \right]|0\rangle, \makebox[2em]{}
\label{eqn:Rh1Rh2} \\
&&\hat{R}_{2}(\lambda)^{\dagger}\hat{R}_{1}(\mu)^{\dagger}|0\rangle=
\sum_{n,m}c_{n\downarrow}^{\dagger}c_{m\uparrow}^{\dagger}
\left[ \theta(n\leq m){\rm e}^{-{\rm i}np(\lambda)}{\rm e}^{-{\rm i}mp(\mu)}
\frac{v(\lambda)-v(\mu)}{v(\lambda)-v(\mu)-{\rm i}U} \right.
\nonumber \\
&&\makebox[2em]{}
\left. +\theta(n> m){\rm e}^{-{\rm i}np(\lambda)}{\rm e}^{-{\rm i}mp(\mu)}
+\theta(n> m){\rm e}^{-{\rm i}mp(\lambda)}{\rm e}^{-{\rm i}np(\mu)}
\frac{{\rm i}U}{v(\lambda)-v(\mu)-{\rm i}U} \right]|0\rangle. \makebox[2em]{}
\label{eqn:Rh2Rh1}
\end{eqnarray}

Note that the two-particle states 
(\ref{eqn:R1R1})-(\ref{eqn:R2R1}) generated by 
$R_{\alpha}(\lambda)^{\dagger}$ are in-states if 
$k(\lambda)<k(\mu)$ and out-states if
$k(\lambda)>k(\mu)$.
Moreover, they are normalized in the sense explained above. 
As for the operators $\hat{R}_{\alpha}(\lambda)^{\dagger}$ 
we observe similar things.
The two-particle states (\ref{eqn:Rh1Rh1})-(\ref{eqn:Rh2Rh1})
are normalized in-states if 
$p(\lambda)>p(\mu)$ and normalized out-states if
$p(\lambda)<p(\mu)$.
These facts,  together with the examples of other integrable 
models~\cite{Tha81,PuWuZh87}
lead us to the following conjecture:
\begin{conjecture}
Provided $k(\lambda_{j})$ is real for $j=1,\cdots,n$, the $n$-particle state 
\begin{equation}
R_{\alpha_{1}}(\lambda_{1})^{\dagger}\cdots
R_{\alpha_{n}}(\lambda_{n})^{\dagger}|0\rangle
\label{eqn:RR0}
\end{equation}
is a normalized in-state if $k(\lambda_{1})<\cdots <k(\lambda_{n})$ and
a normalized out-state if $k(\lambda_{1})>\cdots >k(\lambda_{n})$.\\
Provided $p(\mu_{j})$ is real for $j=1,\cdots,n$, the $n$-particle state 
\begin{equation}
\hat{R}_{\alpha_{1}}(\mu_{1})^{\dagger}\cdots
\hat{R}_{\alpha_{n}}(\mu_{n})^{\dagger}|0\rangle
\label{eqn:RhRh0}
\end{equation}
is a normalized in-state if $p(\mu_{1})>\cdots >p(\mu_{n})$ and
a normalized out-state if $p(\mu_{1})<\cdots <p(\mu_{n})$.
\end{conjecture}
The proof of this conjecture seems difficult for general $n$, 
since 
it seems to be unavoidable to use the series (\ref{eqn:expandtildeT})
and the explicit form (\ref{eqn:tildeL}) of 
$\tilde{{\cal L}}_{m}(\lambda)$.

We have two pairs of normalized one-particle creation operators now, but
as in the case of $B_{a1}(\lambda)$ and $C_{2a}(\lambda)$, 
we do not need to care about both of them in constructing multi-particle 
states.  We may use the operator $R_{\alpha}(\lambda)^{\dagger}$ only
(or $\hat{R}_{\alpha}(\lambda)$ only).
The reason is the 
following. From (\ref{eqn:Roneparticle}) we deduce that 
\begin{equation}
\hat{R}_{\alpha}(\lambda)^{\dagger}|0\rangle=R_{\alpha}
(\tilde{\lambda})^{\dagger}|0\rangle,
\label{eqn:Rh0R0}
\end{equation}
where $p(\lambda)=k(\tilde{\lambda})$.
Hence the action of a mixed product of $R^{\dagger}$ 
and $\hat{R}^{\dagger}$ on the 
vacuum can be expressed in  the form (\ref{eqn:RR0}) by use of 
(\ref{eqn:RdRhd}) and (\ref{eqn:Rh0R0}). 
In particular, one easily obtains
\begin{equation}
\hat{R}_{\alpha_{n}}(\lambda_{n})^{\dagger}\cdots
\hat{R}_{\alpha_{1}}(\lambda_{1})^{\dagger}|0\rangle=
(-1)^{\frac{n(n-1)}{2}}R_{\alpha_{1}}(\tilde{\lambda}_{1})^{\dagger}\cdots
R_{\alpha_{n}}(\tilde{\lambda}_{n})^{\dagger}|0\rangle,
\label{eqn:RhRhRR}
\end{equation}
where $p(\lambda_{j})=k(\tilde{\lambda}_{j})$. The order of the operators 
is reversed when written in terms of $R^{\dagger}_{\alpha}(\lambda)$ instead 
of $\hat{R}^{\dagger}_{\alpha}(\lambda)$.

It turns out that the introduction of the prefactors in 
eqs.~(\ref{eqn:defRd}) and (\ref{eqn:defRhd}) also removes the twist 
(\ref{eqn:twist}) from the commutators of our redefined 
creation and annihilation operators with the Yangian generators. 
Using (\ref{eqn:YB0})-(\ref{eqn:YD}) and some of the formulae in 
\ref{appendix:list}, we obtain
\begin{eqnarray}
&&[Q_{0}^{a}, R_{\alpha}(\lambda)^{\dagger}]=
\frac{1}{2}R_{\beta}(\lambda)^{\dagger}
\sigma_{\beta\alpha}^{a},\label{eqn:Q0R}
\\
&&[Q_{1}^{a}, R_{\alpha}(\lambda)^{\dagger}]=
-\sin k(\lambda)R_{\beta}(\lambda)^{\dagger}
\sigma_{\beta\alpha}^{a}+\frac{U}{2}\varepsilon^{abc}R_{\beta}
(\lambda)^{\dagger}\sigma_{\beta\alpha}^{b}Q_{0}^{c},
\label{eqn:Q1R}
\\
&&[Q_{0}^{a}, \hat{R}_{\alpha}(\lambda)^{\dagger}]=\frac{1}{2}\hat{R}_{\beta}
(\lambda)^{\dagger}
\sigma_{\beta\alpha}^{a},
\label{eqn:Q0Rh} 
\\
&&[Q_{1}^{a}, \hat{R}_{\alpha}(\lambda)^{\dagger}]=
-\sin p(\lambda)\hat{R}_{\beta}
(\lambda)^{\dagger}
\sigma_{\beta\alpha}^{a}-\frac{U}{2}\varepsilon^{abc}\hat{R}_{\beta}
(\lambda)^{\dagger}\sigma_{\beta\alpha}^{b}Q_{0}^{c}.
\label{eqn:Q1Rh}
\end{eqnarray}
These formulae induce an adjoint action of the Yangian on $n$-particle
states~\cite{LeSmi92,MuWa96b}.
Noting that $Q_{0}^{a}|0\rangle=0=Q_{1}^{a}|0\rangle$, 
we obtain the action of the Yangian on the $n=1$ sector as 
\begin{eqnarray}
Q_{0}^{a}R_{\alpha}(\lambda)^{\dagger}|0\rangle&=&
\frac{1}{2}\sigma_{\beta\alpha}^{a}
R_{\beta}(\lambda)^{\dagger}|0\rangle, \\
Q_{1}^{a}R_{\alpha}(\lambda)^{\dagger}|0\rangle&=&
-\sin k(\lambda) \sigma_{\beta\alpha}^{a}
R_{\beta}(\lambda)^{\dagger}|0\rangle.
\end{eqnarray}
Since the action of $Q_{1}^{a}$ is $-2\sin k(\lambda)$ times that of 
$Q_{0}^{a}$, the representation is called the fundamental 
representation $W_{1}(-2\sin k(\lambda))$~\cite{ChaPre,ChaPreBOOK}.

In the two-particle sector ($n=2$) we get 
\begin{eqnarray}
Q_{0}^{a}R_{\alpha}(\lambda_{1})^{\dagger}
R_{\sigma}(\lambda_{2})^{\dagger}
|0\rangle&=&
\left( \frac{1}{2}\sigma_{\beta\alpha}^{a}
\delta_{\rho\sigma}+\frac{1}{2}\delta_{\beta\alpha}
\sigma_{\rho\sigma}^{a}
\right)
R_{\beta}(\lambda_{1})^{\dagger}
R_{\rho}(\lambda_{2})^{\dagger}
|0\rangle, \\
Q_{1}^{a}R_{\alpha}(\lambda_{1})^{\dagger}
R_{\sigma}(\lambda_{2})^{\dagger}
|0\rangle&=&
\left( -\sin k(\lambda_{1})\sigma_{\beta\alpha}^{a}
\delta_{\rho\sigma}-\sin k(\lambda_{2}) \delta_{\beta\alpha}
\sigma_{\rho\sigma}^{a}
+\frac{U}{4}\varepsilon^{abc}\sigma_{\beta\alpha}^{b}
\sigma^{c}_{\rho\sigma}
\right) \nonumber \\
&& \makebox[2em]{} \cdot
R_{\beta}(\lambda_{1})^{\dagger}
R_{\rho}(\lambda_{2})^{\dagger}
|0\rangle. 
\end{eqnarray}
This representation is a tensor product representation 
$W_{1}(-2\sin k(\lambda_{1}))
\otimes W_{1}(-2\sin k(\lambda_{2}))$
with co-multiplication $\Delta$ defined by
\begin{eqnarray}
\Delta(Q_{0}^{a})&=&Q_{0}^{a}\otimes 1 + 1 \otimes Q_{0}^{a}, 
\label{eqn:Delta0}
\\
\Delta(Q_{1}^{a})&=&Q_{1}^{a}\otimes 1 + 1 \otimes Q_{1}^{a}
+U\varepsilon^{abc}Q_{0}^{b}\otimes Q_{0}^{c}.
\label{eqn:Delta1}
\end{eqnarray}
It is four-dimensional and irreducible,
since $k(\lambda_{1})$ and $k(\lambda_{2})$ are real.
Due to the Yangian invariance of the Hamiltonian, these four states 
are degenerate.
Under the sub-algebra su(2) of spins this multiplet is 
decoupled to su(2)-triplet and su(2)-singlet.
We can say that the Yangian Y(su(2)) mixes spin multiplets to 
form a larger multiplet.

Similarly, the $n$-particle states 
$R_{\alpha_{1}}(\lambda_{1})^{\dagger}\cdots
R_{\alpha_{n}}(\lambda_{n})^{\dagger}|0\rangle \ (\alpha_{j}=1,2)$
transform under Y(su(2)) as tensor product representations
$W_{1}(-2\sin k(\lambda_{1}))\otimes\cdots\otimes
W_{1}(-2\sin k(\lambda_{n}))$.
These representations are irreducible under the Yangian 
Y(su(2)), since the quasi-momenta 
$k(\lambda_{j}) $ are real, but not irreducible under the sub-algebra su(2).
In other words, 
these $2^{n}$ states form a multiplet under Y(su(2)), 
while under su(2) they decay into some multiplets according 
to the value of the total spin. 
Thus su(2) is not sufficient to explain the large degeneracy of the 
system in the thermodynamic limit.

The irreducibility leads us to the conclusion that we can construct 
all the $n$-particle states (\ref{eqn:RR0}) out of the 
Yangian highest weight state
\begin{equation}
R_{1}(\lambda_{1})^{\dagger}\cdots
R_{1}(\lambda_{n})^{\dagger}|0\rangle, 
\label{eqn:R1R10}
\end{equation}
by acting with Yangian generators $Q_{n}^{a}$. 
The  wave function of the above state (\ref{eqn:R1R10}) must be 
of plane-wave form, since 
the on-site interaction never occurs between up-spin particles due 
to the Pauli principle. 
Therefore, assuming that the state (\ref{eqn:R1R10}) is a normalized 
one (see Conjecture 1), 
we conjecture that the above 
state (\ref{eqn:R1R10}) is equal to the state
\begin{equation}
c_{\uparrow}^{\dagger}(k(\lambda_{1}))\cdots
c_{\uparrow}^{\dagger}(k(\lambda_{n}))|0\rangle,
\label{eqn:cc0}
\end{equation}
where $c_{\sigma}^{\dagger}(k)=\sum_{j}c_{j\sigma}^{\dagger}
{\rm e}^{-{\rm i}jk}$.
Thus we have got a simple method for constructing multi-particle 
scattering states. They are obtained out of the plane-wave state 
(\ref{eqn:cc0}) by using the Yangian generators (\ref{eqn:Q0}) 
and (\ref{eqn:Q1}). We have already encountered such kind of situation 
in the case of the repulsive $\delta$-function fermi gas~\cite{MuWa96b}.

One can similarly discuss Yangian representations of 
multi-particle states constructed by use of 
$\hat{R}_{\alpha}^{\dagger}(\lambda)$. 
The alert reader will have noticed the different signs in front of $U$
in eqs.~(\ref{eqn:Q1R}) and (\ref{eqn:Q1Rh}), which lead to different 
definitions of the co-multiplication (cf.\ (\ref{eqn:Delta0}),
(\ref{eqn:Delta1})):
\begin{eqnarray}
\Delta'(Q_{0}^{a})&=&Q_{0}^{a}\otimes 1 + 1 \otimes Q_{0}^{a}, 
\label{eqn:Delta'0}
\\
\Delta'(Q_{1}^{a})&=&Q_{1}^{a}\otimes 1 + 1 \otimes Q_{1}^{a}
-U\varepsilon^{abc}Q_{0}^{b}\otimes Q_{0}^{c}.
\label{eqn:Delta'1}
\end{eqnarray}
But this does not cause any contradiction. 
The order of the quasi-momenta in (\ref{eqn:RhRhRR}) is reversed in 
the multi-particle states expressed by 
$R_{\alpha}(\lambda)^{\dagger}$ compared to those expressed by 
$\hat{R}_{\alpha}(\lambda)^{\dagger}$.
This corresponds to the reversed order of the tensor product $\otimes$
in the definition of the 
co-multiplication, which compensates the different sign 
in front of $U$ in (\ref{eqn:Delta1}) and (\ref{eqn:Delta'1}).

The various parameters $U$, $\lambda$, $h$, $v$, $k$ and $p$ 
which we used so far are connected through the formulae 
(\ref{eqn:U4}) and (\ref{eqn:defv}). Thus only two of them are independent. 
As a test of consistency of the results in this section let us 
consider the free fermion limit $U\rightarrow 0$. This limit is 
most conveniently taken for fixed $v$, for if we fix $v=v(\lambda)$ and
$\bar{v}=v(\mu)$ in eqs.\ (\ref{eqn:R1R1})-(\ref{eqn:Rh2Rh1}) and let 
$U$ approach $0$, we see that the products of operators 
$R_{\alpha}(\lambda)^{\dagger}$ and $\hat{R}_{\alpha}(\lambda)^{\dagger}$ 
act like products of creation operators of Bloch states 
$c_{\alpha}^{\dagger}(k_{0})$ on the vacuum. Here $k_{0}=p_{0}$ is 
determined by the corresponding limit in eq.~(\ref{eqn:defv}), 
\begin{equation}
\sin k_{0}=-\frac{v}{2}.
\end{equation}
$\lambda$ and $h$ are now dependent variables. Considering (\ref{eqn:U4}) 
and (\ref{eqn:defv}) for fixed $v$ and small $U$ we find the following 
solutions
\begin{eqnarray}
{\rm i}\cot\lambda&=&1+\frac{U}{4}\left( 1-\frac{v^{2}}{4}
\right)^{-\frac{1}{2}}+O(U^{2}), \label{eqn:U0icotl} \\
{\rm e}^{2h}&=&{\rm i}\left(1-\frac{v^{2}}{4}\right)^{\frac{1}{2}}-
\frac{v}{2}+O(U^{2}).\label{eqn:U0e2h}
\end{eqnarray}
Using these equations and some standard trigonometric identities
we can express all the functions of $h$ and $\lambda$, which enter 
the definition of $\tilde{{\cal L}}_{m}(\lambda)$ (cf.\ (\ref{eqn:tildeL})) 
in terms of $v$ and $U$. Note that (\ref{eqn:U0icotl}) and
(\ref{eqn:U0e2h}) are not the only possible solution of (\ref{eqn:U4}) and 
(\ref{eqn:defv}) for fixed $v$ and small $U$. We choose the branches such 
that $\lim_{U\rightarrow 0}\tilde{{\cal L}}_{m}(\lambda)|_{v}=I_{4}$. 
For small $U$ the odd elements of $\tilde{{\cal L}}_{m}(\lambda)-I_{4}$ 
are of the order of $U^{\frac{1}{2}}$ and the even elements are of the 
order of $U$. Thus only the first sum on the rhs of (\ref{eqn:expandtildeT}) 
contributes in order $U^{\frac{1}{2}}$ to the odd elements of 
$\tilde{{\cal T}}(\lambda)-I_{4}$, and we obtain 
\begin{eqnarray}
C_{2\alpha}(\lambda)&=&{\rm i}^{\alpha-3}\frac{{\rm e}^{-h}}{\cos\lambda}
\sum_{m}c_{m\alpha}^{\dagger}{\rm e}^{-{\rm i}mk_{0}}+O(U^{\frac{3}{2}}),
\\
B_{3-\alpha,1}(\lambda)&=&{\rm i}^{1-\alpha}\frac{{\rm e}^{h}}{\cos\lambda}
\sum_{m}c_{m\alpha}^{\dagger}{\rm e}^{-{\rm i}mp_{0}}+O(U^{\frac{3}{2}}),
\end{eqnarray}
where ${\rm e}^{\pm h}/ \cos\lambda=O(U^{\frac{1}{2}})$. 
Since $D_{\beta\beta}(\lambda)=1+O(U)$ $(\beta=1,2)$, it follows from the 
definitions (\ref{eqn:defRd}) and (\ref{eqn:defRhd}) that 
\begin{equation}
\lim_{U\rightarrow 0}R_{\alpha}(\lambda)^{\dagger}|_{v}=
\lim_{U\rightarrow 0}\hat{R}_{\alpha}(\lambda)^{\dagger}|_{v}=
c^{\dagger}_{\alpha}(k_{0}).
\end{equation}
The corresponding formulae for $R_{\alpha}(\lambda)$ and 
$\hat{R}_{\alpha}(\lambda)$ are true by hermitian conjugation. 
Eqs.\ (\ref{eqn:RdRd})-(\ref{eqn:RRhd}) turn into the usual anticommutators 
between fermi operators, since
\begin{equation}
\lim_{U\rightarrow 0} r(\lambda,\mu)|_{v,\bar{v}}={\cal P}.
\end{equation}
We see that we may interpret the Zamolodchikov-Faddeev algebra as a 
deformation of the anticommutators between fermi operators with deformation 
parameter $U$.

\subsection{Bound States}

One of the delicate points in Bethe ansatz calculations is the 
question of completeness. The completeness of the 
coordinate Bethe ansatz for the Hubbard model under periodic 
boundary conditions was discussed by E{\ss}ler, Korepin and
Schoutens~\cite{EKS}. They showed that the Bethe wave functions are 
highest weight with respect to the su(2)$\oplus$su(2) symmetry generated 
by $S^{a}$ and $\eta^{a}$~\cite{EKS}. Hence, each solution of the 
Bethe ansatz equations correspond to a su(2)$\oplus$su(2) multiplet. 
They counted the number of Bethe ansatz solutions assuming Takahashi's 
string hypothesis~\cite{Ta72} 
to be valid and multiplied by the multiplicities of
the su(2)$\oplus$su(2) multiplets. The resulting number is equal to 
the dimension of the Hilbert space. Note however that, although 
a matter of common belief now, the string hypothesis is waiting 
for a proof since 25 years.

How to pose the question of completeness in our infinite chain formalism?
As we have seen in the preceding section the operators 
$R_{\alpha}(\lambda)^{\dagger}$ and $\hat{R}_{\alpha}(\lambda)^{\dagger}$ 
create single electrons in scattering states. The operator $D_{21}(\lambda)$ 
creates a bound pair of electrons. As we have learned in section 5, 
there are no operators that create more than two particles 
among the elements of the monodromy matrix.  As we will see, however, 
the string hypothesis suggests the existence of bound states of pairs. 
How to define the corresponding bound state operators?

To obtain a guess, let us recall the string hypothesis. In the 
following we will denote the spin rapidities of the coordinate Bethe ansatz
by $\Lambda_{j}$ and the momenta by $k_{j}$ (for details cf.\ \cite{Ta72}). 
According to the string hypothesis there are two types of 
string solutions of the Bethe ansatz equations. 
\begin{enumerate}
\item
($\Lambda$-string) 
$m \ \Lambda_{j}$'s form a string configuration, in which 
the real parts of the $\Lambda_{j}$'s are the same while 
the imaginary parts are arranged at equal spacing of ${\rm i}U/2$.
The center of the string should be real.  
\item
($k$-$\Lambda$-string)
$2m \ \k_{i}$'s and $m \ \Lambda_{j}$'s form a string configuration.
The values of $k_{i}$'s and $\Lambda_{j}$'s are 
\begin{eqnarray*}
&&k_{1}=\pi-\arcsin(\Lambda'+{\rm i}mU/4), \\
&&k_{2}=\arcsin(\Lambda'+{\rm i}(m-2)U/4), \\
&&k_{3}=\pi-k_{2} \\
&& \vdots \\
&&k_{2m-2}=\arcsin(\Lambda'-{\rm i}(m-2)U/4), \\
&&k_{2m-1}=\pi-k_{2m-2} \\
&&k_{2m}=\pi-\arcsin(\Lambda'-{\rm i}mU/4), \\
&&\Lambda_{j}=\Lambda'+{\rm i}(m+1-2j)U/4, \ \Lambda' \ {\rm real}, \ \ \ 
j=1,2,\cdots, m.
\end{eqnarray*}
\end{enumerate}
These solutions should be exact in the thermodynamic limit. Since 
we are dealing with the zero density vacuum, there should be no 
spin excitations, and we do not have to consider the $\Lambda$-string 
here.

We wish to obtain an operator which creates a $2m$-$k$-$\Lambda$-string
(for short we shall simply call it as ``$2m$-string").
To begin with, we shall deal with the 2-string ($m=1$ case). 
There are 2 particles involved, one with spin up and the other one with 
spin down. The wave function of the 2-string state is of the form 
\begin{equation}
\sum_{m,n} c_{m \uparrow}^{\dagger} c_{n \downarrow}^{\dagger}
\left\{
\theta(m\geq n){\rm e}^{- {\rm i}(mk + np)}+
\theta(m< n){\rm e}^{- {\rm i}(nk + mp)}
\right\} |0\rangle,
\end{equation}
where $\sin k-\sin p={\rm i}U/2$ and $k+p$ real.
Therefore, by comparison with eq.~(\ref{eqn:D21v}), 
the 2-string state is proportional to 
$D_{21}(\lambda)|0\rangle$ with an appropriate choice of $\lambda$. 
It can be easily seen by explicit use of (\ref{eqn:expandtildeT}) and 
(\ref{eqn:tildeL}) that it is also proportional to 
$C_{22}(\lambda')C_{21}(\lambda^{\prime\prime})|0\rangle$, if 
$\lambda'$ and 
$\lambda^{\prime\prime}$ satisfy the following conditions;
\begin{eqnarray}
&&p(\lambda')=\pi-k(\lambda^{\prime\prime})
 \ \ {\rm mod} \ 2\pi, \label{eqn:cond1}\\
&&k(\lambda^{\prime\prime})=p(\lambda) 
\ \ {\rm mod} \ 2\pi, \label{eqn:cond2}\\
&&k(\lambda')=k(\lambda) \ \ {\rm mod} \ 2\pi. \label{eqn:cond3}
\end{eqnarray}
These are three conditions 
for three parameters $\lambda$, $\lambda'$, $\lambda^{\prime\prime}$, 
which at first sight seems to violate the arbitrariness of $\lambda$. 
Yet there is a redundancy in these equations. 
(\ref{eqn:cond1}) and (\ref{eqn:cond2}) imply
\begin{equation}
p(\lambda')=\pi-p(\lambda) \ \ {\rm mod} \ 2\pi,
\end{equation}
which is compatible with (\ref{eqn:cond3}) by taking into account 
the constraint (\ref{eqn:sinksinp}).
Thus we have obtained two possible 2-string creation operators, which 
are connected with each other by 
\begin{eqnarray}
&& D_{21}(\lambda)|0\rangle =
\frac{{\rm i}{\rm e}^{h(\lambda')+h(\lambda^{\prime\prime})}
\cos\lambda^{\prime\prime}\cos^{2}\lambda' \sin\lambda'}{
\cos^{2}\lambda} \nonumber \\
&&\makebox[3em]{}
\cdot
\frac{(1-{\rm e}^{{\rm i}(p(\lambda')-k(\lambda^{\prime\prime}))})
(1-{\rm e}^{{\rm i}(k(\lambda^{\prime\prime})-k(\lambda'))})}{1-
{\rm e}^{{\rm i}(p(\lambda')
-k(\lambda'))}}
C_{22}(\lambda')C_{21}(\lambda^{\prime\prime})|0\rangle.
\label{eqn:DC22C21}
\end{eqnarray}
$\lambda$, $\lambda'$ and $\lambda^{\prime\prime}$ in this equation have to
satisfy (\ref{eqn:cond1})-(\ref{eqn:cond3}). 
Note that it follows from (\ref{eqn:C2aC2b}) that
\begin{equation}
C_{22}(\lambda')C_{21}(\lambda^{\prime\prime})=
-C_{21}(\lambda')C_{22}(\lambda^{\prime\prime}).
\end{equation}

Let us proceed with  the general $2m$-string states. 
We conjecture that the creation operator of a $2m$-$k$-$\Lambda$-string 
can be expressed as 
\begin{equation}
C^{(2m)}_{2}(\lambda_{1},\cdots,\lambda_{2m})=
C_{22}(\lambda_{1})C_{21}(\lambda_{2})
C_{22}(\lambda_{3})C_{21}(\lambda_{4})\cdots
C_{22}(\lambda_{2m-1})C_{21}(\lambda_{2m}),
\end{equation}
where 
\begin{eqnarray}
&& k(\lambda_{2s})+p(\lambda_{2s-1})=\pi  \ ({\rm mod} \  2\pi),\\
&&\sin k(\lambda_{2s-1})=\sin k(\lambda_{1})+\frac{{\rm i}U(s-1)}{2},
\ (s=1,\cdots,m).
\end{eqnarray}
Following previous works~\cite{bound,SoWa83,KuSk,KuSm} 
we shall call this operator bound-state operator. 
The expression on the rhs is a formal one and should be 
interpreted as a ``composite operator" (see \ref{appendix:comp}).
One can easily verify that the functions
$\sin k(\lambda_i)$ form the same configuration as in the
$k$-$\Lambda$-string, if their center 
\begin{equation}
\zeta=\frac{1}{2m}\sum_{i=1}^{2m}\sin k(\lambda_{i})
=\sin k(\lambda_{1})+\frac{{\rm i}U(m-2)}{4}
\end{equation}
is real.

We can normalize the bound-state operator by a method similar
to that in the case of the scattering states created by
$R_{\alpha}(\la)^{\dagger}$.
Let
\begin{equation}
D^{(2m)}_{22}(\lambda_{1},\cdots,\lambda_{2m})=
D_{22}(\lambda_{1})D_{22}(\lambda_{2})
D_{22}(\lambda_{3})D_{22}(\lambda_{4})\cdots
D_{22}(\lambda_{2m-1})D_{22}(\lambda_{2m}).
\end{equation}
We define a normalized bound-state operator as
\begin{equation}
  R^{(2m)}(\lambda_{1},\cdots,\lambda_{2m})^{\dagger}=
C^{(2m)}_{2}(\lambda_{1},\cdots,\lambda_{2m})
D^{(2m)}_{22}(\lambda_{1},\cdots,\lambda_{2m})^{-1}.
\end{equation}
Similar definitions of bound state operators have appeared before in
the literature. They have been applied to the
XXZ-chain~\cite{KuSk,KuSm} and to the
attractive $\delta$-function gas~\cite{bound,NLSattractive}.
Note that it is not a priori clear, if we can apply the commutation
rules for the elements of the monodromy matrix to obtain the
commutators of $2m$-string operators (cf.\ \ref{appendix:comp}). However,
if we assume we can, we get the following reasonable result,
\begin{eqnarray}
&&R^{(2m)}(\lambda_{i})^{\dagger}R^{(2n)}(\mu_{j})^{\dagger}=
\frac{\zeta-\eta+(n+m){\rm i}U/4}{\zeta-\eta-(n+m){\rm i}U/4}
\ \frac{\zeta-\eta+|n-m|{\rm i}U/4}{\zeta-\eta-|n-m|{\rm i}U/4}\nonumber \\
&& \makebox[2em]{}\cdot
\prod_{s=1}^{\min(m,n)-1}\left[
\frac{\zeta-\eta+(n+m-2s){\rm i}U/4}{\zeta-\eta-(n+m-2s){\rm i}U/4}
\right]^{2}R^{(2n)}(\mu_{j})^{\dagger}
R^{(2m)}(\lambda_{i})^{\dagger},
\label{eqn:R2mR2n}\\
&&R^{(2m)}(\lambda_{i})^{\dagger}R_{a}(\mu)^{\dagger}=
\frac{\zeta-\sin k(\mu)+{\rm i}Um/4}{\zeta-\sin k(\mu)-{\rm i}Um/4}
R_{a}(\mu)^{\dagger}R^{(2m)}(\lambda_{i})^{\dagger},
\label{eqn:R2mR}
\end{eqnarray}
where
$\zeta$ is the center 
of the $2m$-string and $\eta$ is the
center of the $2n$-string. The factor on the rhs of (\ref{eqn:R2mR2n})  
is the $S$-matrix between a $2m$-string and a $2n$-string, 
and that on the rhs of (\ref{eqn:R2mR}) is the $S$-matrix between 
a $2m$-string and a particle.
(\ref{eqn:R2mR2n}) is of the same form as the $S$-matrix for the 
scattering of bound states of magnons in the XXX-chain~\cite{KuSm}.

As for the transformation under the Yangian Y(su(2))
we can easily show that 
\begin{equation}
[Q_{0}^{a}, R^{(2m)}(\lambda_{i})^{\dagger}]=0=
[Q_{1}^{a}, R^{(2m)}(\lambda_{i})^{\dagger}],
\label{eqn:QaR2m}
\end{equation}
which follows from eq.\ (\ref{eqn:QaC22C21}) and from 
the commutativity of $Q_{n}^{a}$ and $D_{22}(\lambda_{j})$. From 
(\ref{eqn:QaR2m})
the action of the Yangian Y(su(2)) on a $2m$-string state 
can be derived as 
\begin{equation}
Q_{0}^{a}R^{(2m)}(\lambda_{i})^{\dagger}|0\rangle=0, \ \ 
Q_{1}^{a}R^{(2m)}(\lambda_{i})^{\dagger}|0\rangle=0,  
\end{equation}
i.e.\ the $2m$-string state is singlet under Y(su(2)).

\section{Concluding Remarks and Discussion}
\setcounter{equation}{0}

We have developed the QISM for the Hubbard model on 
the infinite interval with respect to the zero density vacuum. 
The $R$-matrix (\ref{eqn:deftildeR}) thus obtained
is greatly simplified in comparison with the $R$-matrix 
of the finite periodic model. 
In particular, it reveals a hidden rational structure, 
which arises from a certain combination of the functions $\rho_{i}$ in 
eq.~(\ref{eqn:matR}). This structure was discovered earlier 
by Ramos and Martins~\cite{RaMa96} as part of the exchange relation 
for the Hubbard model on the finite interval. 
Along with the simplified 
$R$-matrix we obtained the 
asymptotic expansion (\ref{eqn:expandA}) of the 
submatrix $A(\lambda)$ of the 
monodromy matrix, which naturally provides a representation 
of Y(su(2)) 
and generates an infinite series of mutually commuting 
Yangian invariant operators, which is including the Hamiltonian.
We thus clarified the origin of the Yangian symmetry of the Hubbard
model.

We constructed creation and annihilation operators of elementary
excitations. There are two types of excitations over the zero
density vacuum, one-particle excitations and $2m$-string excitations.
We showed that they are distinguished by their behavior under the action
of the Yangian. All $2m$-string excitations are Yangian singlets, whereas
the $n$-fold one-particle excitations transform like $n$-fold tensor
products of the fundamental Yangian representation.

The interaction of two elementary excitations is described by their
$S$-matrix, which is given by the commutation relations of the 
corresponding operators. We calculated these commutation relations and
thus the $S$-matrix. The $2m$-string excitations are Yangian singlet,
which means that they have no internal structure. Therefore their mutual
interaction produces only phase shifts, their $S$-matrix (6.58)
is a scalar factor. In contrast, the scattering of one-particle
excitations involves the spin. The $S$-matrix is a $4 \times 4$-matrix.
Up to a factor of minus one, which accounts for the fermionic nature
of the excitations, it agrees with the rational submatrix
$r(\lambda,\mu)$ of the $R$-matrix in the thermodynamic limit. We
obtained two alternative pairs of one-particle creation operators.
The corresponding $S$-matrices are reciprocal to one-another, which
means that, if one of the pairs creates in-states, the other one
creates out-states. Our creation and annihilation operators combine
into a left and right representation of the Zamolodchikov-Faddeev algebra.
Although it should be obvious, we would like to emphasize that our
one-particle operators in equation (6.6)-(6.11) are not abstract
quantities, but explicitly given in terms of the elementary creation
and annihilation operators of Wannier states. In the limit of vanishing
coupling ($U \rightarrow 0$) our one-particle operators turn into
creation and annihilation operators of Bloch states. The
Zamolodchikov-Faddeev algebra can thus be understood as a deformation
of the anticommutators between fermi operators with deformation
parameter $U$.

There is a known standard method for the solution of the ``quantum inverse 
problem"~\cite{Tha81}, which consists in expressing the fermi 
operators for electrons in Wannier states $c_{j\sigma}, \ 
c_{j\sigma}^{\dagger}$ in terms of quasi-particle operators 
$R_{\alpha}(\lambda)$,  
$R_{\alpha}(\lambda)^{\dagger}$ and $R^{(2m)}(\lambda)$, 
$R^{(2m)}(\lambda)^{\dagger}$. 
In analogy to the classical inverse scattering method, we have to derive 
a quantum Gelfand-Levitan equation. To this end we should investigate 
the analytic properties of the monodromy matrix. Due to 
the complicated structure of the $R$-matrix, this may be a 
difficult task. We leave it for future work. The solution of the 
inverse problem will enable us to calculate Green's function with 
respect to the considered vacuum, as it was done before for other 
integrable models~\cite{CrThWi80,KuSk,KuSm}.

So far our work has been limited to the four cases of 
uncorrelated vacua, for which the up-spin or down-spin 
orbitals are either completely filled or vacant.
Unfortunately, these cases are not of particular interest
from the point of view of condensed matter physics. Correlation
functions are rather trivial and may be obtained directly within
the coordinate Bethe ansatz solution of Lieb and Wu. Hence there
is a clear need to develop a method for the algebraic construction
of excitations over a {\it correlated} vacuum. In a first step
we had to renormalize ${\cal T}_{mn} (\lambda)$ with respect to
this vacuum. The properties of the vacuum enter the formalism,
they are not results of it, and therefore have to be obtained
by independent means. Assume we had done this step. Then the next
question would be, if the vacuum is a Fock vacuum with respect
to some of the entries of the renormalized monodromy matrix. This
is, of course, not clear a priori, but is a necessary requirement
for an algebraic construction of eigenstates based on commutation
relations.

The development of a version of the QISM for interacting vacua is
certainly a hard and challenging task, and it may be more appropriate
to start with a more simple model than the Hubbard model. It seems
however, that such kind of method would finally lead to a complete
understanding of the Hubbard model, too. We would like to consider our
work as a first necessary step towards this goal.

For a further investigation of the Hubbard model by algebraic means
the physically interesting half filled case seems to be most
appropriate. At half filling the $2m$-strings disappear from the
spectrum \cite{EsKo94}. There are only two pairs of {\it independent}
one-particle excitations, which have been described as ``spinons''
and ``holons'' \cite{EsKo94}. In analogy to the results in section
6.1 we expect the space of states for a given set of rapidities to
be spanned by the action of Y(su(2))$\oplus$Y(su(2)) Yangian generators
on Yangian highest weight states. We further know that the half filled
ground state (at zero magnetic field) is singlet with respect to
Y(su(2))$\oplus$Y(su(2)) \cite{UgKo94}. This suggests that form factors
may be calculated by purely algebraic means \cite{JiMiBOOK}.

\section*{Acknowledgements}
We would like to thank M.~Wadati for his interest in this work, 
for his hospitality and for bringing us together. We benefited 
from stimulating discussions with M.~J.~Martins, 
M.~Shiroishi, H.~Frahm and J.~Suzuki. 
S.~M. is grateful to N.~Nagaosa for his encouragement. 
The final preparation of this manuscript during a 
visit of F.~G. at the University of Tokyo became 
possible through financial support by the DFG (grant number 
Go 825/1-1). In Bayreuth F.~G. is postdoc fellow of the 
``Graduiertenkolleg nichtlineare Spektroskopie und Dynamik".
He likes to express his gratitude to 
F.~G.~Mertens and D.~Haarer for their support.

\setcounter{section}{0}
\renewcommand{\thesection}{Appendix \Alph{section}}

\section{Singular Terms in the Infinite-chain Formalism}
\label{appendix:singular}
\renewcommand{\theequation}{\Alph{section}.\arabic{equation}}
\setcounter{equation}{0}

In this appendix we calculate the limits
\begin{eqnarray}
  (U_{+}(\lambda,\mu)^{-1})_{\alpha\beta,\gamma\delta}
&=&\lim_{n\rightarrow\infty}
   (U_{n}(\lambda,\mu)^{-1})_{\alpha\beta,\gamma\delta}, \label{eqn:Uplus}\\
  U_{-}(\lambda,\mu)_{\alpha\beta,\gamma\delta}&=&\lim_{n\rightarrow -\infty}
   U_{n}(\lambda,\mu)_{\alpha\beta,\gamma\delta}, \label{eqn:Uminus}
\end{eqnarray}
which determine the elements of $\tilde{{\cal T}}(\lambda)\otimes
\tilde{{\cal T}}(\mu)$ via equation (\ref{eqn:TTUTU}).
$U_{n}(\lambda,\mu)^{-1}$ and $U_{n}(\lambda,\mu)$ are defined by 
eq.(\ref{eqn:defU}), where $V(\lambda)$ is given according 
to (\ref{eqn:defV}).  $V^{(2)}(\lambda,\mu)$ is easily obtained
by direct calculation. Its diagonal consists of the elements of 
$V(\lambda)\otimes_{{\rm s}}V(\mu)$. Due to normal ordering there 
appear additional non-vanishing off-diagonal elements, 
\begin{eqnarray}
&&V^{(2)}(\lambda,\mu)_{12,21}=V^{(2)}(\lambda,\mu)_{13,31}=
-{\rm i}\sin\lambda\sin\mu, \\
&&V^{(2)}(\lambda,\mu)_{14,23}=-V^{(2)}(\lambda,\mu)_{14,32}=
-{\rm i}\sin\lambda \ \cos\mu, \\
&&V^{(2)}(\lambda,\mu)_{24,42}=V^{(2)}(\lambda,\mu)_{34,43}=
{\rm i}\cos\lambda\cos\mu, \\
&&V^{(2)}(\lambda,\mu)_{23,41}=-V^{(2)}(\lambda,\mu)_{32,41}=
-{\rm i}\cos\lambda\sin\mu, \\
&&V^{(2)}(\lambda,\mu)_{14,41}=-{\rm e}^{h(\lambda)+h(\mu)}.
\end{eqnarray}
Note that $V^{(2)}(\lambda,\mu)$ is upper triangular. Since the 
diagonals of $V^{(2)}(\lambda,\mu)$ and $V(\lambda)\otimes_{{\rm s}}
V(\mu)$ agree, $V^{(2)}(\lambda,\mu)$ can be diagonalized by an upper 
triangular matrix $U(\lambda,\mu)$ whose diagonal elements are all 
unity, and
\begin{equation}
  V^{(2)}(\lambda,\mu)=U(\lambda,\mu)(
V(\lambda)\otimes_{{\rm s}}
V(\mu))U(\lambda,\mu)^{-1}.
\label{eqn:VUVVU}
\end{equation}
It turns out that the non-vanishing off-diagonal elements of 
$U(\lambda,\mu)$ are simple rational functions of the Boltzmann 
weights $\rho_{j}=\rho_{j}(\lambda,\mu)$. They are obtained as
\begin{eqnarray}
&& U(\lambda,\mu)_{12,21}=
U(\lambda,\mu)_{13,31}=\frac{-{\rm i}\rho_{2}}{\rho_{10}}, \\
&& U(\lambda,\mu)_{14,23}=
-U(\lambda,\mu)_{14,32}=\frac{{\rm i}\rho_{6}}{\rho_{3}-\rho_{1}}, \\
&&U(\lambda,\mu)_{24,42}=
U(\lambda,\mu)_{34,43}=\frac{{\rm i}\rho_{2}}{\rho_{9}}, \\
&& U(\lambda,\mu)_{23,41}=
-U(\lambda,\mu)_{32,41}=\frac{{\rm i}\rho_{6}}{\rho_{5}-\rho_{4}},\\
&&U(\lambda,\mu)_{14,41}=\frac{-\rho_{5}}{\rho_{5}-\rho_{4}}.
\end{eqnarray}
Before proceeding further let us introduce some shorthand notation. 
Instead of $f(\lambda)$, $g(\mu)$ we will write $f$, $\bar{g}$. 
The bar means that the argument of the function if $\mu$.
Using this convention 
and equation (\ref{eqn:VUVVU}) we obtain
\begin{eqnarray}
  U_{n}(\lambda,\mu)&=&U(\lambda,\mu)
(V^{-n}\otimes_{{\rm s}}\bar{V}^{-n})U(\lambda,\mu)^{-1}(V^{n}
\otimes_{{\rm s}}\bar{V}^{n}), \label{eqn:Um} \\
  U_{n}(\lambda,\mu)^{-1}&=&
(V^{-n}\otimes_{{\rm s}}\bar{V}^{-n})U(\lambda,\mu)(V^{n}
\otimes_{{\rm s}}\bar{V}^{n})U(\lambda,\mu)^{-1}, \label{eqn:Un1}
\end{eqnarray}
Comparing (\ref{eqn:Um}), (\ref{eqn:Un1}) and
(\ref{eqn:Uplus}), (\ref{eqn:Uminus}) we find that we have 
to calculate the following limits
\begin{eqnarray}
&&l_{1}^{\pm}=\lim_{n\rightarrow\pm\infty}\frac{{\rm i}\rho_{2}}{\rho_{10}}
( 1-{\rm e}^{{\rm i}n(p-\bar{p})}), \label{eqn:l1} \\
&&l_{2}^{\pm}=\lim_{n\rightarrow\pm\infty}\frac{-{\rm i}\rho_{2}}{\rho_{9}}
( 1-{\rm e}^{{\rm i}n(k-\bar{k})}), \label{eqn:l2} \\
&&l_{3}^{\pm}=\lim_{n\rightarrow\pm\infty}\frac{{\rm i}\rho_{6}}{\rho_{1}-
\rho_{3}}
( 1-{\rm e}^{{\rm i}n(p-\bar{k})}), \label{eqn:l3}\\
&&l_{4}^{\pm}=\lim_{n\rightarrow\pm\infty}\frac{{\rm i}\rho_{6}}{\rho_{4}-
\rho_{5}}
( 1-{\rm e}^{{\rm i}n(k-\bar{p})}), \label{eqn:l4}\\
&&l_{5}^{+}=\lim_{n\rightarrow \infty}
\left\{
\frac{\rho_{3}}{\rho_{1}-\rho_{3}}+
\frac{2\rho_{6}^{2}}{(\rho_{1}-\rho_{3})(\rho_{4}-\rho_{5})}
{\rm e}^{{\rm i}n(p-\bar{k})}\right. \nonumber \\
&&\makebox[2em]{}
+\left.\frac{\rho_{5}}{\rho_{4}-\rho_{5}}
{\rm e}^{{\rm i}n(k+p-\bar{k}-\bar{p})}\right\}, \label{eqn:l5}\\
&&l_{6}^{-}=\lim_{n\rightarrow -\infty}
\left\{
\frac{\rho_{5}}{\rho_{4}-\rho_{5}}+
\frac{2\rho_{6}^{2}}{(\rho_{1}-\rho_{3})(\rho_{4}-\rho_{5})}
{\rm e}^{{\rm i}n(k-\bar{p})}\right. \nonumber \\
&&\makebox[2em]{}
+\left.\frac{\rho_{3}}{\rho_{1}-\rho_{3}}
{\rm e}^{{\rm i}n(k+p-\bar{k}-\bar{p})}\right\}, \label{eqn:l6}
\end{eqnarray}
These limits exist in the sense of generalized functions. 
It turns out that they all can be reduced to the following formula
\begin{equation}
\lim_{n\rightarrow\pm\infty}\frac{1-{\rm e}^{{\rm i}(p-\bar{p})n}}{
{\rm e}^{{\rm i}\bar{p}}-{\rm e}^{{\rm i}p}}=
\frac{1}{{\rm e}^{{\rm i}\bar{p}}-{\rm e}^{{\rm i}(p\pm{\rm i}\varepsilon)}},
\label{eqn:singularformula}
\end{equation}
where $\varepsilon$ is an 
infinitesimal positive number. 
The limits exist in the domain $\pm\Im(p-\bar{p})\geq 0$ 
A proof of the above formula can be
obtained in two steps. First show by acting on a test function that 
\begin{equation}
\lim_{n\rightarrow\pm\infty}{\rm p.v.}
\frac{{\rm e}^{{\rm i}(p-\bar{p})n}}{
{\rm e}^{{\rm i}(p-\bar{p})}-1}=
\pm\pi\tilde{\delta}(p-\bar{p}), 
\end{equation}
where $\tilde{\delta}$ is the periodic $\delta$-function
\begin{equation}
  \tilde{\delta}(p)=\sum_{n=-\infty}^{\infty}\delta(p-2n\pi).
\end{equation}
Then use a periodic version of the Plemelj formula
(cf. page 501 of ref.~\cite{FaTa})
\begin{equation}
\frac{1}{1-{\rm e}^{{\rm i}(p\pm{\rm i}\varepsilon)}}=
{\rm p.v.}
\frac{1}{1-{\rm e}^{{\rm i}p}}\pm\pi\tilde{\delta}(p),
\label{eqn:SoPl}
\end{equation}
to obtain (\ref{eqn:singularformula}).

Using (\ref{eqn:singularformula}) we find, for example, 
\begin{eqnarray}
l_{1}^{\pm}&=&
-\frac{{\rm i}{\rm e}^{-h-\bar{h}}}{\sin\lambda\sin\mu}
\lim_{n\rightarrow\pm\infty}\frac{1-{\rm e}^{{\rm i}(p-\bar{p})n}}
{{\rm e}^{{\rm i}\bar{p}}-{\rm e}^{{\rm i}p}} \nonumber\\
&=&
-\frac{{\rm i}{\rm e}^{-h-\bar{h}}}{\sin\lambda\sin\mu}
\frac{1}{{\rm e}^{{\rm i}\bar{p}}-{\rm e}^{{\rm i}(p\pm{\rm
      i}\varepsilon)}},\label{eqn:l1r}
\end{eqnarray}
and similarly
\begin{eqnarray}
l_{2}^{\pm}&=&
\frac{{\rm i}{\rm e}^{h+\bar{h}}}{\sin\lambda\sin\mu}
\frac{1}{{\rm e}^{{\rm i}\bar{k}}-{\rm e}^{{\rm i}(k\pm{\rm i}
\varepsilon)}}, \label{eqn:l2r} \\
l_{3}^{\pm}&=&
\frac{{\rm i}{\rm e}^{-h+\bar{h}}}{\sin\lambda\sin\mu}
\frac{1}{{\rm e}^{{\rm i}\bar{k}}-{\rm e}^{{\rm i}(p\pm{\rm i}
\varepsilon)}},\label{eqn:l3r} \\
l_{4}^{\pm}&=&
\frac{{\rm i}{\rm e}^{h-\bar{h}}}{\sin\lambda\sin\mu}
\frac{1}{{\rm e}^{{\rm i}\bar{p}}-{\rm e}^{{\rm i}(k\pm{\rm i}
\varepsilon)}}.\label{eqn:l4r}
\end{eqnarray}
Note that the prefactors on the rhs of the above formulae when 
expressed in terms 
of $k$, $\bar{k}$, $p$, $\bar{p}$ are all regular at the singularities of 
the second factors. The remaining limits are obtained as 
\begin{eqnarray}
l_{5}^{+}&=&
\frac{
(1+{\rm e}^{{\rm i}(\bar{k}+\bar{p})})(1+{\rm e}^{{\rm i}(k+p)})}
{{\rm e}^{{\rm i}\bar{p}}-{\rm e}^{{\rm i}k}}
\left\{
\frac{{\rm e}^{{\rm i}\bar{p}}+{\rm e}^{{\rm i}k}}{
{\rm e}^{{\rm i}(\bar{k}+\bar{p})}-{\rm e}^{{\rm i}(k+p+
{\rm i}\varepsilon)}}
-\frac{2}{{\rm e}^{{\rm i}\bar{k}}-{\rm e}^{{\rm i}(p+
{\rm i}\varepsilon)}}
\right\}, \label{eqn:l5r}\\
l_{6}^{-}&=&
\frac{
(1+{\rm e}^{{\rm i}(\bar{k}+\bar{p})})(1+{\rm e}^{{\rm i}(k+p)})}
{{\rm e}^{{\rm i}\bar{k}}-{\rm e}^{{\rm i}p}}
\left\{
\frac{{\rm e}^{{\rm i}\bar{k}}+{\rm e}^{{\rm i}p}}{
{\rm e}^{{\rm i}(\bar{k}+\bar{p})}-{\rm e}^{{\rm i}(k+p-
{\rm i}\varepsilon)}}
-\frac{2}{{\rm e}^{{\rm i}\bar{p}}-{\rm e}^{{\rm i}(k-
{\rm i}\varepsilon)}}
\right\}, \label{eqn:l6r}
\end{eqnarray}
The last two equations require consideration of the consistency of the
occurring singularities.

In particular, in case of (\ref{eqn:l5r}) 
the following two conditions must be satisfied, (i)
$\Im(p)>\Im(\bar{k})$ if $k+p$ and $\bar{k}+\bar{p}$ are real, 
(ii) $\Im(k)>\Im(\bar{p})$, if $\bar{k}$ and $p$ are real. 
These two conditions have to be consistent with the constraint 
(\ref{eqn:sinksinp}).

Consider the first condition. We can use equation (\ref{eqn:constr}) to 
obtain
\begin{equation}
\Im( p-\bar{k})=-{\rm arcsinh}
\left( \frac{U}{4\cos((k+p)/2)}\right)-{\rm arcsinh}
\left( \frac{U}{4\cos((\bar{k}+\bar{p})/2)}\right).
\end{equation}
Hence $\Im(p-\bar{k})$ is positive for positive $U$, if 
$\frac{\pi}{2}<|\frac{k+p}{2}|<\pi$, and positive for negative $U$, if 
$|\frac{k+p}{2}|<\frac{\pi}{2}$. Note that the same restrictions on the 
parameters were obtained in section 5.1 as conditions on $D_{21}(\lambda)$ 
to create a bound state. 

Consider the second condition above. The constraint (\ref{eqn:sinksinp}) 
has two branches of solutions for $p$ as a function of $k$, which are fixed 
by choosing ${\rm sgn}(\Im(p))$. We may therefore choose 
$\Im(k)>0>\Im(\bar{p})$, and (ii) will be satisfied. 
Moreover, this choice of branch ensures the prefactor 
$({\rm e}^{{\rm i}\bar{p}}-{\rm e}^{{\rm i}k})^{-1}$ in (\ref{eqn:l5r}) 
to be nonsingular as $k$ approaches $p$ on the real axis.

Equation (\ref{eqn:l6r}) may be discussed in a similar way. Now the 
two conditions become (i)
$\Im(k)<\Im(\bar{p})$ if $k+p$ and $\bar{k}+\bar{p}$ are real, and
(ii) $\Im(p)>\Im(\bar{k})$, if $\bar{p}$ and $k$ are real. 
Condition (i) implies the same restriction as above, 
$\frac{\pi}{2}<|\frac{k+p}{2}|<\pi$ for $U>0$, and 
$|\frac{k+p}{2}|<\frac{\pi}{2}$ for $U<0$. 
(ii) can again be satisfied by an appropriate choice of branch of $p$ as 
a function of $k$. 

We are now in a position to consider the weak limits
\begin{equation}
(\tilde{{\cal T}}(\lambda)\otimes_{s}\tilde{{\cal T}}
(\mu))_{\alpha\beta,\gamma\delta}=
(U_{+}(\lambda,\mu)^{-1})_{\alpha\beta,\epsilon\varphi}
\tilde{{\cal T}}^{(2)}(\lambda,\mu)_{\epsilon\varphi,\eta\rho}
U_{-}(\lambda,\mu)_{\eta\rho,\gamma\delta}.
\label{eqn:tildeTtildeT}
\end{equation}
We have to check, whether the functions on the rhs of this equation 
exist on a common domain. Clearly, this has to be done equation 
by equation. We find the following 16 combinations of functions 
$l_{j}^{\pm}$:
\begin{eqnarray*}
&&\{ l_{1}^{+},l_{1}^{-} \}, \  
\{ l_{1}^{+},l_{3}^{-} \}, \  
\{ l_{3}^{+},l_{5}^{+},l_{1}^{-} \}, \  
\{ l_{3}^{+},l_{5}^{+},l_{3}^{-} \}, \  
\{ l_{1}^{+},l_{4}^{-},l_{6}^{-} \},   
\\
&&\{ l_{1}^{+},l_{2}^{-} \}, \    
\{ l_{3}^{+},l_{5}^{+},l_{4}^{-},l_{6}^{-} \}, \  
\{ l_{3}^{+},l_{5}^{+},l_{2}^{-} \}, \  
\{ l_{4}^{+},l_{1}^{-} \}, \  
\{l_{4}^{+},l_{3}^{-}\}, 
\\
&&\{ l_{2}^{+},l_{1}^{-} \}, \  
\{ l_{2}^{+},l_{3}^{-} \}, \  
\{ l_{4}^{+},l_{4}^{-},l_{6}^{-} \}, \  
\{ l_{4}^{+},l_{2}^{-} \}, \  
\{ l_{2}^{+},l_{4}^{-},l_{6}^{-} \}, \ 
\{l_{2}^{+},l_{2}^{-}\}.
\end{eqnarray*}
It turns out that each of these combinations is compatible. However, 
they are not compatible all together.

Now we should solve (\ref{eqn:tildeTtildeT}) for 
$\tilde{{\cal T}}^{(2)}(\lambda,\mu)_{\alpha\beta,\gamma\delta}$ and 
insert the result into (\ref{eqn:exchangeT2}) to 
obtain the commutation relations between the elements of the monodromy 
matrix. We should do that equation by equation and should take care of 
the compatibility of the domains of the occurring functions. This 
would be a cumbersome task and would, moreover, obstruct the 
algebraic structure of our problem. Therefore we leave 
mathematical rigor at this point and proceed more formally. 

First note that there occur products of the form $l_{1}^{+}l_{1}^{-}$ 
on the rhs of (\ref{eqn:tildeTtildeT}). These products 
are not well defined, since the regularization requires two 
limits, and, when acting on a test function, the result will 
depend on the order of these limits. ``This indicates the highly singular 
operator character"\footnote{We are citing Sklyanin~\cite{Sk}.} 
of some of the elements of the 
monodromy matrix. In the following we 
will therefore exclude the singular points of the
functions $l_{j}^{\pm}$ (\ref{eqn:l1r}) - (\ref{eqn:l6r}). We will assume 
that $k\neq\bar{k}$ for $k$, $\bar k$ real etc. 
Then we can omit the regularizations 
$\pm{\rm i}\varepsilon$ in $l_{j}^{\pm}$, and 
\begin{equation}
U_{+}(\lambda,\mu)^{-1}=U(\lambda,\mu)^{-1}, \ \ 
U_{-}(\lambda,\mu)=U(\lambda,\mu).
\label{eqn:UpUUmU}
\end{equation}
We insert this result into (\ref{eqn:tildeTtildeT}), treat 
(\ref{eqn:tildeTtildeT}) as a matrix equation and solve for 
$\tilde{{\cal T}}^{(2)}(\lambda,\mu)$, 
\begin{equation}
\tilde{{\cal T}}^{(2)}(\lambda,\mu)=U(\lambda,\mu)(
\tilde{{\cal T}}(\lambda)\otimes_{s}\tilde{{\cal T}}(\mu))
U(\lambda,\mu)^{-1}.
\end{equation}
Now (\ref{eqn:exchangeT2}) implies (\ref{eqn:newRTT}), and 
the $R$-matrix (\ref{eqn:deftildeR}) is obtained by using
(\ref{eqn:UpUUmU}) in eq.(\ref{eqn:tildeRpm}).

\section{List of Commutation Rules}
\label{appendix:list}
\setcounter{equation}{0}

\subsection{Elementary Commutators}

In this appendix we provide a complete list of the commutation 
rules encoded in the exchange relation (\ref{eqn:newRTT}) in terms of the 
submatrices $A(\lambda),\cdots,D(\lambda)$ of the 
monodromy matrix $\tilde{{\cal T}}(\lambda)$. As mentioned in 
section \ref{sec:passage}
these submatrices generate sub-algebras of (\ref{eqn:newRTT}),
\begin{eqnarray}
r(\lambda,\mu)(A(\lambda)\otimes A(\mu))&=&
(A(\mu)\otimes A(\lambda))r(\lambda,\mu), \label{eqn:AA}\\
s(\lambda,\mu)(D(\lambda)\otimes D(\mu))&=&
(D(\mu)\otimes D(\lambda))s(\lambda,\mu), \label{eqn:DD}
\end{eqnarray}
\begin{eqnarray}
&&\frac{\rho_{4}(\lambda,\mu)}{\rho_{1}(\lambda,\mu)}r(\lambda,\mu)
{B_{1a}(\lambda)\choose B_{2a}(\lambda)}
\otimes
{ B_{1a}(\mu)\choose B_{2a}(\mu)}
\nonumber \\
&&\makebox[4em]{}
=
{ B_{1a}(\mu)\choose B_{2a}(\mu)}
\otimes
{B_{1a}(\lambda)\choose B_{2a}(\lambda)}
, \label{eqn:BaBa}\\
&&\frac{\rho_{4}(\lambda,\mu)}
{\rho_{1}(\lambda,\mu)-\rho_{3}(\lambda,\mu)}r(\lambda,\mu)
{B_{11}(\lambda)\choose B_{21}(\lambda)}
\otimes
{ B_{12}(\mu)\choose B_{22}(\mu)}
\nonumber \\
&& \makebox[4em]{}
=
{B_{12}(\mu)\choose B_{22}(\mu)}
\otimes
{B_{11}(\lambda)\choose B_{21}(\lambda)}
, \label{eqn:B1B2}
\end{eqnarray}
\begin{eqnarray}
&&(C_{a1}(\lambda),C_{a2}(\lambda)) \otimes
(C_{a1}(\mu),C_{a2}(\mu)) \nonumber \\
&&\makebox[1em]{}=
(C_{a1}(\mu),C_{a2}(\mu)) \otimes
(C_{a1}(\lambda),C_{a2}(\lambda))
\frac{\rho_{4}(\lambda,\mu)}{\rho_{1}(\lambda,\mu)}r(\lambda,\mu)
, \label{eqn:CaCa}\\
&&(C_{11}(\lambda),C_{12}(\lambda)) \otimes
(C_{21}(\mu),C_{22}(\mu)) \nonumber \\
&&\makebox[1em]{}=
(C_{21}(\mu),C_{22}(\mu)) \otimes
(C_{11}(\lambda),C_{12}(\lambda))
\frac{\rho_{4}(\lambda,\mu)}{\rho_{1}(\lambda,\mu)-\rho_{3}
(\lambda,\mu)}r(\lambda,\mu).
 \label{eqn:C1C2}
\end{eqnarray}
The matrix $s(\lambda,\mu)$ in (\ref{eqn:DD}) is defined as 
\begin{equation}
  s(\lambda,\mu)={\rm diag}
\left(1,\frac{\rho_{4}}{\rho_{4}-\rho_{5}},
\frac{\rho_{1}-\rho_{3}}{\rho_{1}},1 \right){\cal P}.
\end{equation}
The commutators of $B(\lambda)$,$C(\lambda)$ and $D(\lambda)$ 
with the submatrix $A(\mu)$ are given as 
\begin{eqnarray}
\label{eqn:BA1}
r(\lambda, \mu) \,
\left( {B_{11} (\lambda) \choose B_{21} (\lambda)} \otimes A (\mu) \right)
& = & \frac{{\rm i}\r_{10} (\lambda, \mu)}{\r_4 (\lambda, \mu)} \,
A(\mu) \otimes {B_{11} (\lambda) \choose B_{21} (\lambda)} , \\
\label{eqn:BA2}
r(\lambda, \mu) \,
\left( {B_{12} (\lambda) \choose B_{22} (\lambda)} \otimes A (\mu) \right)
& = & - \frac{{\rm i}\r_1 (\lambda, \mu)}{\r_9 (\lambda, \mu)} \,
A(\mu) \otimes {B_{12} (\lambda) \choose B_{22} (\lambda)} , \\
\label{eqn:CA1}
\frac{{\rm i}\r_{10} 
(\lambda, \mu)}{\r_4 (\lambda, \mu)} \, (C_{11} (\lambda), C_{12} (\lambda))
\otimes A (\mu) & = & (A(\mu) \otimes (C_{11} (\lambda), C_{12} (\lambda))
      \, r(\lambda, \mu) , \\
     \label{eqn:CA2}
- \frac{{\rm i}\r_1 (\lambda, \mu)}{\r_9 (\lambda, \mu)} \, 
(C_{21} (\lambda), C_{22} (\lambda))
\otimes A (\mu) & = & (A(\mu) \otimes (C_{21} (\lambda), C_{22} (\lambda))
\, r(\lambda, \mu) .
\end{eqnarray}
\begin{eqnarray}
\label{eqn:DAg}
[D_{11} (\lambda), A(\mu)] & = & [D_{22} (\lambda), A(\mu)]  \,\, = \,\, 0
, \\
\label{eqn:DA1}
\, D_{12} (\lambda) A (\mu) &=&
-\frac{\r_1 (\lambda, \mu)}{\r_9 (\lambda, \mu)} 
\frac{\r_{4} (\lambda, \mu)}{\r_{10} (\lambda, \mu)}\, A(\mu) D_{12}(\lambda) 
, \\
\label{eqn:DA2}\, 
D_{21} (\lambda) A (\mu) &=&-
\frac{\r_9 (\lambda, \mu)}{\r_1 (\lambda, \mu)} 
\frac{\r_{10} (\lambda, \mu)}{\r_4 (\lambda, \mu)} \, A (\mu) D_{21} (\lambda)
.
\end{eqnarray}
The commutators between the entries of $B(\lambda)$ and 
$C(\mu)$ are
\begin{eqnarray}
B_{a1}(\lambda)C_{1b}(\mu)&=&
-\frac{\rho_{10}(\lambda,\mu)^{2}}{\rho_{1}
(\lambda,\mu)\rho_{4}(\lambda,\mu)}C_{1b}(\mu)B_{a1}(\lambda), 
\label{eqn:B1C1} \\
B_{a1}(\lambda)C_{2b}(\mu)&=&
\frac{\rho_{10}(\lambda,\mu)}{\rho_{9}
(\lambda,\mu)}C_{2b}(\mu)B_{a1}(\lambda), 
\label{eqn:B1C2} \\
B_{a2}(\lambda)C_{1b}(\mu)&=&
\frac{\rho_{10}(\lambda,\mu)}{\rho_{9}
(\lambda,\mu)}C_{1b}(\mu)B_{a2}(\lambda), 
\label{eqn:B2C1} \\
B_{a2}(\lambda)C_{2b}(\mu)&=&
-\frac{\rho_{1}(\lambda,\mu)\rho_{4}(\lambda,\mu)}{\rho_{9}
(\lambda,\mu)^{2}}C_{2b}(\mu)B_{a2}(\lambda). \label{eqn:B2C2}
\end{eqnarray}
Finally, there are the following commutators between $B(\lambda)$, 
$C(\lambda)$ and the submatrix $D(\lambda)$,
\begin{eqnarray}
B_{a1}(\lambda)D(\mu)&=&{\rm i}
\left(
\begin{array}{cc}
\frac{\rho_{10}}{\rho_{4}} & 0\\
0 & -\frac{\rho_{1}}{\rho_{9}}
\end{array}
\right)
D(\mu)
\left(
\begin{array}{cc}
1 & 0 \\
0 & \frac{\rho_{1}-\rho_{3}}{\rho_{1}}
\end{array}
\right)
B_{a1}(\lambda), \label{eqn:B1D}\\
B_{a2}(\lambda)D(\mu)&=&{\rm i}
\left(
\begin{array}{cc}
\frac{\rho_{10}}{\rho_{4}} & 0 \\
0 & -\frac{\rho_{1}}{\rho_{9}}
\end{array}
\right)
D(\mu)
\left(
\begin{array}{cc}
\frac{\rho_{4}}{\rho_{4}-\rho_{5}} & 0 \\
0 & 1
\end{array}
\right)
B_{a2}(\lambda), \label{eqn:B2D}\\
C_{1a}(\lambda)D(\mu)&=&-{\rm i}
\left(
\begin{array}{cc}
1 & 0 \\
0 & \frac{\rho_{1}}{\rho_{1}-\rho_{3}}
\end{array}
\right)
D(\mu)
\left(
\begin{array}{cc}
\frac{\rho_{4}}{\rho_{10}} & 0 \\
0 & -\frac{\rho_{9}}{\rho_{1}}
\end{array}
\right)
C_{1a}(\lambda), \label{eqn:C1D}\\
C_{2a}(\lambda)D(\mu)&=&-{\rm i}
\left(
\begin{array}{cc}
\frac{\rho_{4}-\rho_{5}}{\rho_{4}} & 0 \\
0 & 1
\end{array}
\right)
D(\mu)
\left(
\begin{array}{cc}
\frac{\rho_{4}}{\rho_{10}} & 0 \\
0 & -\frac{\rho_{9}}{\rho_{1}}
\end{array}
\right)
C_{2a}(\lambda). \label{eqn:C2D}
\end{eqnarray}

\subsection{Commutators with the quantum determinant}

Eqs. (\ref{eqn:BA1})-(\ref{eqn:DA2}) imply the following commutators 
of the quantum determinant 
${\rm Det}_{q}(A(\mu))$ with the remaining entries of the 
monodromy matrix
\begin{eqnarray}
             {\rm Det}_{q} (A(\mu)) B_{a1} (\lambda) &=&
 -\frac {\r_4 (\lambda, \mu) \r_4 (\lambda, \check{\mu})}
 {\r_{10} (\lambda, \mu) \r_{10} (\lambda, \check{\mu})}            
 \frac{v(\lambda)-v(\mu)}{v(\lambda)-v(\mu) + {\rm i}U} \, B_{a1} (\lambda) 
{\rm Det}_{q} (A(\mu))  
 , \ \label{eqn:detB1} \\
      {\rm Det}_{q} (A(\mu)) B_{a2} (\lambda) &=&
 - \frac{\r_9 (\lambda, \mu) \r_9 (\lambda, \check{\mu})}
 {\r_1 (\lambda, \mu) \r_1 (\lambda, \check{\mu})}
 \frac{v(\lambda)-v(\mu)}{v(\lambda)-v(\mu) + {\rm i}U} \, B_{a2} (\lambda) 
{\rm Det}_{q} (A(\mu)) 
 ,\  \label{eqn:detB2}
\end{eqnarray}
\begin{eqnarray}
     \label{eqn:detC1}
      {\rm Det}_{q} (A(\mu)) C_{1a} (\lambda) & = &
      -\frac{\r_{10} (\lambda, \mu) \r_{10} (\lambda, \check{\mu})}
             {\r_4 (\lambda, \mu) \r_4 (\lambda, \check{\mu})}
     \frac{v(\lambda)-v(\mu) + {\rm i}U}{v(\lambda)-v(\mu)}
     \, C_{1a} (\lambda) {\rm Det}_{q} (A(\mu)) , \ \\
     \label{eqn:detC2}
      {\rm Det}_{q} (A(\mu)) C_{2a} (\lambda)& = &
      - \frac{\r_1 (\lambda, \mu) \r_1 (\lambda, \check{\mu})}
             {\r_9 (\lambda, \mu) \r_9 (\lambda, \check{\mu})}
     \frac{v(\lambda)-v(\mu) + {\rm i}U}{v(\lambda)-v(\mu)} \, 
C_{2a} (\lambda) {\rm Det}_{q} (A(\mu))  .  \ 
\end{eqnarray}
\begin{eqnarray}
      &&{\rm Det}_{q} (A(\mu)) D_{12} (\lambda)= \nn \\
      &&\makebox[2em]{} \frac{\r_9 (\lambda, \mu) \r_9 (\lambda, \check{\mu})}
           {\r_1 (\lambda, \mu) \r_1 (\lambda, \check{\mu})}
      \frac{\r_{10} (\lambda, \mu) \r_{10} (\lambda, \check{\mu})}
           {\r_4 (\lambda, \mu) \r_4 (\lambda, \check{\mu})}
       D_{12} (\lambda) {\rm Det}_{q} (A(\mu))           
       \label{eqn:detD1}  ,  \\
      &&{\rm Det}_{q} (A(\mu)) D_{21} (\lambda) = \nn \\
      &&\makebox[2em]{}\frac{\r_1 (\lambda, \mu) \r_1 (\lambda, \check{\mu})}
           {\r_9 (\lambda, \mu) \r_9 (\lambda, \check{\mu})}
      \frac{\r_4 (\lambda, \mu) \r_4 (\lambda, \check{\mu})}
           {\r_{10} (\lambda, \mu) \r_{10} (\lambda, \check{\mu})}
      D_{21} (\lambda) {\rm Det}_{q} (A(\mu)) ,     \label{eqn:detD2}  \\
      && [{\rm Det}_{q} (A(\mu)), D_{11}(\lambda)]=0=[{\rm Det}_{q}
(A(\mu)), D_{22}(\lambda)],
        \label{eqn:detD3}
\end{eqnarray}
where $\check{\mu}$ is defined by $v(\check{\mu})=v(\mu)-{\rm i}U$.

\section{Composite operators}
\label{appendix:comp}
\setcounter{equation}{0}

For the construction of bound state operators corresponding to the 
$2m$-string in section 6.2, we have to introduce composite operators, 
which are formal products of entries of the  monodromy matrix.
Our definition of the 2-string creation operator, for instance, is 
\begin{equation}
R^{(2)}(\lambda_{1},\lambda_{2})^{\dagger}=
C_{22}(\lambda_{1})C_{21}(\lambda_{2})D_{22}(\lambda_{2})^{-1}
D_{22}(\lambda_{1})^{-1},
\label{eqn:R2}
\end{equation}
where 
\begin{equation}
p(\lambda_{1})+k(\lambda_{2})=\pi \ ({\rm mod} \ 2\pi).
\label{eqn:pl1kl2}
\end{equation}
Because of the constraint (\ref{eqn:sinksinp}), this implies 
\begin{equation}
\sin k(\lambda_{1})-\sin k(\lambda_{2})=\frac{{\rm i}U}{2}.
\end{equation}
Thus $k_{1}$ and $k_{2}$ cannot both be real, and it follows from 
(\ref{eqn:C21v}) and (\ref{eqn:C22v}) that $C_{22}(\lambda_{1})$ 
and $C_{21}(\lambda_{2})$ in the products on the rhs of (\ref{eqn:R2}) 
cannot both create bounded states. They are not simultaneously physical 
operators. Despite this fact, we can provide {\it the product \/} 
$C_{22}(\lambda_{1})C_{21}(\lambda_{2})$ with physical meaning, when 
$\sin k(\lambda_{1})-{\rm i}U/4$ is real.
Eq.\ (\ref{eqn:tildeTtildeT}) implies 
\begin{equation}
     C_{22} (\lambda_{1}) C_{21} (\lambda_{2}) =
     l_{3}^{-} \tilde{{\cal T}}^{(2)}(\lambda_{1}, \lambda_{2})_{44,14} +
\tilde{{\cal T}}^{(2)}(\lambda_{1}, \lambda_{2})_{44,32}.
\label{eqn:C22C21}
\end{equation}
The matrix elements $\tilde{{\cal T}}^{(2)} (\lambda_{1},
\lambda_{2})_{\alpha\beta, \gamma\delta}$ are well-defined
through a series representation similar to (\ref{eqn:expandtildeT}). 
We may take
(\ref{eqn:C22C21}) as a definition of the composite operator on the left
hand side. Its domain is the domain of $l_{3}^{-}$. It is determined
by the condition $\Im (p(\lambda_{1})) \leq \Im (k(\lambda_{2}))$, which
leads to the same discussion as above eq.\ (\ref{eqn:constr}) or below 
eq.\ (\ref{eqn:l6r}).

In principle, we could iterate the renormalization procedure explained 
in section 3 and in \ref{appendix:singular}. We could define 
${\cal L}_{m}^{(k)}(\lambda_{1},\cdots,\lambda_{m})$ as the $k$-fold 
graded tensor product of $L$-matrices at site $m$ and could introduce its 
expectation value $V^{(k)}(\lambda_{1},\cdots,\lambda_{m})$ 
which governs the renormalization of the $k$-fold tensor product of 
monodromy matrices ${\cal T}_{mn}^{(k)}(\lambda_{1},\cdots,\lambda_{k})$.
We would obtain a renormalized tensor product 
$\tilde{{\cal T}}^{(k)}(\lambda_{1},\cdots,\lambda_{k})$ and the 
commutation relations between the entries of 
$\tilde{{\cal T}}^{(k)}(\lambda_{1},\cdots,\lambda_{k})$ 
and 
$\tilde{{\cal T}}^{(l)}(\mu_{1},\cdots,\mu_{l})$ 
(let $\tilde{{\cal T}}^{(1)}(\lambda)=\tilde{{\cal T}}(\lambda)$).
We guess that such kind of procedure would solve the completeness problem 
in a satisfactory way. Unfortunately, it seems to be practically impossible 
to do these calculation, because of the 
increasing dimension of the involved matrices. 

A composite operator is not a mere product of the original operators. 
So it is not obvious whether or not the commutation rules between composite 
operators follow from iterating the commutation rules of its factors, 
as obtained from (\ref{eqn:newRTT}). Nevertheless, we assumed so and 
investigated some of the consequences of this assumption. This way 
we obtained the $S$-matrices (\ref{eqn:R2mR2n}) and (\ref{eqn:R2mR}) for 
composite operators, which look very reasonable. Another consequence of 
such kind of formal procedure is that 
all $2m$-string states are Yangian singlet. In case of the two-string 
this can be seen as follows.
Using (\ref{eqn:CaCa}) with 
$\rho_{4}(\lambda_{2},\lambda_{1})\neq 0$,   
$\rho_{1}(\lambda_{2},\lambda_{1})=0$ and
$r(\lambda_{2},\lambda_{1})=(1+{\cal P})/2$, we get
\begin{equation}
C_{2\alpha}(\lambda_{1})C_{2\beta}(\lambda_{2})=
-C_{2\beta}(\lambda_{1})C_{2\alpha}(\lambda_{2}).
\label{eqn:C2aC2b}
\end{equation}
\makebox[0em]{}From this equation one can derive the following formula
\begin{equation}
[Q_{0}^{a}, C_{22}(\lambda_{1})C_{21}(\lambda_{2})]=0=
[Q_{1}^{a}, C_{22}(\lambda_{1})C_{21}(\lambda_{2})],
\label{eqn:QaC22C21}
\end{equation}
which indicates that the composite operator 
$C_{22}(\lambda_{1})C_{21}(\lambda_{2})$  creates a 
Yangian invariant pair of up- and down-spin particles.

\section{Conserved Quantities of the Hubbard Model}
\label{appendix:cons}
\renewcommand{\theequation}{\Alph{section}.\arabic{equation}}
\setcounter{equation}{0}

Explicit expressions for higher conserved quantities have been derived by 
several authors~\cite{Gro,GraMa}. In contrast to the case of the nonlinear 
Schr{\"o}dinger model these quantities are not in one to one 
correspondence to the conserved quantities of the free fermion model. 
Let us define creation operators of Bloch states as 
\begin{equation}
c_{\sigma}^{\dagger}(k)=\sum_{j}c_{j\sigma}^{\dagger}
{\rm e}^{-{\rm i}jk}, \ \ \ k\in  [-\pi,\pi].
\end{equation}
In terms of these operators we can define two sequences of operators
\begin{eqnarray}
F_{n}&=&-\frac{1}{\pi}
\int_{-\pi}^{\pi}dk\  \sin\left(n\left(k+\frac{\pi}{2}
\right)\right)c_{\sigma}^{\dagger}(k)c_{\sigma}(k), \\
\tilde{F}_{n}&=&-\frac{1}{\pi}
\int_{-\pi}^{\pi}dk\  \cos\left(n\left(k+\frac{\pi}{2}
\right)\right)c_{\sigma}^{\dagger}(k)c_{\sigma}(k). 
\end{eqnarray}
All of these operators are mutually commuting, and we have 
$\hat{H}(U=0)=F_{1}$. Using the results of ref.~\cite{GraMa}
we obtain the following higher conserved quantities.
\begin{eqnarray}
&&H_{1}=\hat{H},
\\
&&H_{2}=-{\rm i}\sum_{j,\sigma}(c_{j+2,\sigma}^{\dagger}c_{j,\sigma}-
c_{j,\sigma}^{\dagger}c_{j+2,\sigma})\nonumber \\
&&\makebox[3em]{}+{\rm i}U\sum_{j,\sigma}
[(c_{j+1,\sigma}^{\dagger}c_{j,\sigma}-c_{j,\sigma}^{\dagger}c_{j+1,\sigma})
(n_{j,-\sigma}+n_{j+1,-\sigma}-1)],\\
&&H_{3}=\sum_{j,\sigma}(c_{j+3,\sigma}^{\dagger}c_{j,\sigma}+
c_{j,\sigma}^{\dagger}c_{j+3,\sigma})\nonumber \\
&&\makebox[3em]{}-U\sum_{j,\sigma}
\left[
(c_{j+2,\sigma}^{\dagger}c_{j,\sigma}+c_{j,\sigma}^{\dagger}c_{j+2,\sigma})
(n_{j,-\sigma}+n_{j+1,-\sigma}+n_{j+2,-\sigma}-\frac{3}{2})
\right.
\nonumber \\
&&\makebox[4em]{}
+(c_{j+2,\sigma}^{\dagger}c_{j+1,\sigma}-c_{j+1,\sigma}^{\dagger}
c_{j+2,\sigma})
(c_{j+1,-\sigma}^{\dagger}c_{j,-\sigma}-c_{j,-\sigma}^{\dagger}
c_{j+1,-\sigma})
\nonumber \\
&&\makebox[4em]{}
+\frac{1}{2}(c_{j+1,\sigma}^{\dagger}c_{j,\sigma}-c_{j,\sigma}^{\dagger}
c_{j+1,\sigma})
(c_{j+1,-\sigma}^{\dagger}c_{j,-\sigma}-c_{j,\sigma}^{\dagger}
c_{j+1,-\sigma})
\nonumber \\
&&\makebox[4em]{}
\left. -\left( n_{j+1,\sigma}-\frac{1}{2}\right)
\left( n_{j,-\sigma}-\frac{1}{2}\right)
-\frac{1}{2}\left( n_{j,\sigma}-\frac{1}{2}\right)
\left( n_{j,-\sigma}-\frac{1}{2}\right)+\frac{3}{8}
\right]
\nonumber \\
&&\makebox[4em]{}
+U^{2}\sum_{j,\sigma}
(c_{j+1,\sigma}^{\dagger}c_{j,\sigma}+c_{j,\sigma}^{\dagger}c_{j+1,\sigma})
\left[ \left( n_{j+1,-\sigma}-\frac{1}{2}\right)
\left( n_{j,-\sigma}-\frac{1}{2}\right)+\frac{1}{4}
\right],\\
&&H_{4}={\rm i}\sum_{j,\sigma}(c_{j+4,\sigma}^{\dagger}c_{j,\sigma}-
c_{j,\sigma}^{\dagger}c_{j+4,\sigma})\nonumber \\
&&\makebox[3em]{}-{\rm i}U\sum_{j,\sigma}
[(c_{j+3,\sigma}^{\dagger}c_{j,\sigma}-c_{j,\sigma}^{\dagger}c_{j+3,\sigma})
(n_{j,-\sigma}+n_{j+1,-\sigma}+n_{j+2,-\sigma}+n_{j+3,-\sigma}-2)
\nonumber \\
&&\makebox[4em]{}
+(c_{j+2,\sigma}^{\dagger}c_{j,\sigma}+c_{j,\sigma}^{\dagger}
c_{j+2,\sigma})\nonumber \\
&&\makebox[6em]{}
\cdot(
c_{j,-\sigma}^{\dagger}c_{j-1,-\sigma}-c_{j-1,-\sigma}^{\dagger}
c_{j,-\sigma}+
c_{j+1,-\sigma}^{\dagger}c_{j,-\sigma}-c_{j,-\sigma}^{\dagger}
c_{j+1,-\sigma}+
\nonumber \\
&&\makebox[6em]{}
c_{j+2,-\sigma}^{\dagger}c_{j+1,-\sigma}-c_{j+1,-\sigma}^{\dagger}
c_{j+2,-\sigma}+
c_{j+3,-\sigma}^{\dagger}c_{j+2,-\sigma}-c_{j+2,-\sigma}^{\dagger}
c_{j+3,-\sigma}
)
\nonumber \\
&&\makebox[4em]{}
-(c_{j+1,\sigma}^{\dagger}c_{j,\sigma}-c_{j,\sigma}^{\dagger}c_{j+1,\sigma})
(n_{j-1,-\sigma}+n_{j,-\sigma}+n_{j+1,-\sigma}+n_{j+2,-\sigma}-2)]
\nonumber \\
&&\makebox[4em]{}
+{\rm i}U^{2}\sum_{j,\sigma}
\left\{
\ 
(c_{j+2,\sigma}^{\dagger}c_{j,\sigma}-c_{j,\sigma}^{\dagger}c_{j+2,\sigma})
\left[ 
\left( n_{j+1,-\sigma}-\frac{1}{2}\right)
\left( n_{j,-\sigma}-\frac{1}{2}\right)\right.\right.
\nonumber \\
&&\makebox[5em]{}
+\left.\left( n_{j+2,-\sigma}-\frac{1}{2}\right)
\left( n_{j,-\sigma}-\frac{1}{2}\right)+
\left( n_{j+2,-\sigma}-\frac{1}{2}\right)
\left( n_{j+1,-\sigma}-\frac{1}{2}\right)
\right] \nonumber \\
&&\makebox[4em]{}
+(c_{j+1,\sigma}^{\dagger}c_{j,\sigma}+c_{j,\sigma}^{\dagger}c_{j+1,\sigma})
\left[
(c_{j,-\sigma}^{\dagger}c_{j-1,-\sigma}+c_{j-1,-\sigma}^{\dagger}c_{j,-\sigma})
\left(n_{j+1,-\sigma}-\frac{1}{2}\right)+\right.
\nonumber \\
&&\makebox[5em]{}
\left.\left.
+(c_{j+2,-\sigma}^{\dagger}c_{j+1,-\sigma}+c_{j+1,-\sigma}^{\dagger}
c_{j+2,-\sigma})
\left(n_{j,-\sigma}-\frac{1}{2}\right)\right]\ \right\}
\nonumber \\
&&\makebox[4em]{}
+{\rm i}U^{3}\sum_{j,\sigma}\frac{1}{4}
(c_{j+1,\sigma}^{\dagger}c_{j,\sigma}-c_{j,\sigma}^{\dagger}c_{j+1,\sigma})
(n_{j,-\sigma}+n_{j+1,-\sigma}-1).
\end{eqnarray}
These operators have the following properties, 
(i) they are Hermitian and renormalized, $H_{s}|0\rangle=0$, (ii) they are 
SO(4) invariant, $[H_{s},S^{a}]=0=[H_{s},\eta^{a}]$, (iii) $H_{s}(U=0)=
F_{s}$, (iv) their action on the one-particle states 
$|k,\sigma\rangle=c_{\sigma}^{\dagger}(k)|0\rangle$ is obtained as 
\begin{eqnarray}
H_{1}|k,\sigma\rangle&=&\left[ 
-2\cos k-\frac{U}{2}\right]|k,\sigma\rangle, \\
H_{2}|k,\sigma\rangle&=&\left[ 
2\sin 2k+2U\sin k\right]|k,\sigma\rangle, \\
H_{3}|k,\sigma\rangle&=&\left[ 
2\cos 3k-\frac{3U}{2}+3U\cos 2k+U^{2}\cos k \right]|k,\sigma\rangle, \\
H_{4}|k,\sigma\rangle&=&\left[ 
-2\sin 4k+U(4\sin k-4\sin 3k)-\frac{3U^{2}}{2}\sin 2k +
\frac{U^{3}}{2}\sin k \right]|k,\sigma\rangle. 
\end{eqnarray}
The latter formulae lead us to the conjecture (\ref{eqn:lndetA}) in section 
5.3. We make a remark here that there is an arbitrariness in the 
definition of higher conserved quantities by adding linear combinations 
of lower ones. We do not know what is the most natural choice for  
higher conserved quantities.

\end{document}